\documentclass[12pt,draftclsnofoot,onecolumn]{IEEEtran}

\makeatletter
\newcommand\semihuge{\@setfontsize\semihuge{22.3}{22}}
\makeatother

\usepackage{algpseudocode}
\usepackage{algorithm}
\usepackage{algorithmicx}

\usepackage{lipsum} 
\usepackage{slashbox}
\usepackage[dvips]{color}
\usepackage{comment}
\usepackage{todonotes}
\usepackage{epsf}
\usepackage{epsfig}
\usepackage{times}
\usepackage{epsfig}
\usepackage{cite}
\usepackage{graphicx}
\usepackage{mathtools}
\usepackage{mathrsfs}
\usepackage{amssymb}
\usepackage{pdfpages} 
\usepackage{epstopdf}
\usepackage{float}
\newfloat{algorithm}{t}{lop}
\usepackage{subfig}

\usepackage{dsfont}
\usepackage{lettrine} 
\usepackage{amsmath,epsfig,amssymb,algorithm,algpseudocode,amsthm,cite,url}
\usepackage{caption}
\usepackage{bbm}
\allowdisplaybreaks
\usepackage{csquotes}
\newcommand{\R}{\ensuremath{{\mathbb R}}}
\newcommand{\E}{\ensuremath{{\mathbb E}}}
\usepackage{multirow}
\usepackage{verbatim}
\usepackage{subfig}
\usepackage[english]{babel}
\usepackage{amsmath,amssymb}

\usepackage{verbatim}

\newtheorem{corollary}{Corollary}
\newtheorem{theorem}{\bf Theorem}

\newtheorem{lemma}{\bf Lemma}

\newtheorem{remark}{Remark}

\begin{document}
\bstctlcite{IEEEexample:BSTcontrol}

\title{\semihuge  Spatial Motifs for Device-to-Device Network Analysis in Cellular Networks \vspace{-0.3cm}}    

\author{\IEEEauthorblockN{  Tengchan Zeng\IEEEauthorrefmark{1}, Omid Semiari\IEEEauthorrefmark{2}, Walid Saad\IEEEauthorrefmark{1}, and My T. Thai\IEEEauthorrefmark{3}}\vspace{-0.05cm}\\
	\IEEEauthorblockA{
		\small \IEEEauthorrefmark{1}Wireless@VT, Electrical and Computer Engineering Department, Virginia Tech, VA, USA,\\
		\IEEEauthorrefmark{2}Department of Electrical and Computer Engineering, Georgia Southern University, Statesboro, GA, USA,\\
		\IEEEauthorrefmark{3}Department of Computer and Information Science and Engineering, University
		of Florida, Gainesville, FL, USA, \\	Emails: \IEEEauthorrefmark{1}\{tengchan, walids\}@vt.edu, \IEEEauthorrefmark{2}osemiari@georgiasouthern.edu, \IEEEauthorrefmark{3}mythai@cise.ufl.edu.\\
		\thanks{\textcolor{black}{A preliminary version of this work was presented in part at the IEEE Asilomar Conference on Signals, Systems, and Computers, Pacific Grove, CA, USA, Oct. 2017 \cite{zeng2017interference}}. This research was supported by the U.S. National Science Foundation under Grant CNS-1513697. }
	}
\vspace{-0.92cm}}
\maketitle
\vspace{-0.8cm}
\vspace{-0.1cm}
\begin{abstract}
\textcolor{black}{
	Device-to-device (D2D) communication is a promising
	approach to efficiently disseminate critical or viral information. 
	Reaping the benefits of D2D-enabled networks is contingent upon choosing the optimal content dissemination policy subject to resource and user distribution constraints.
	In this paper, a novel D2D network analysis framework is proposed to study the impacts of frequently occurring subgraphs, known as motifs, on D2D network performance and to determine an effective content dissemination strategy.
	In the proposed framework, the distribution of devices in the D2D network is modeled as a Thomas cluster process (TCP), and two graph structures, the star and chain motifs, are studied in the communication graph.  
	Based on the properties of the TCP, closed-form analytical expressions for the statistical significance, the outage probability, as well as the average throughput per device, are derived. 
	Simulation results corroborate the analytical derivations and show the influence of different system topologies on the occurrence of motifs and the D2D system throughput.
	More importantly, the results highlight that, as the statistical significance of motifs increases, the system throughput will initially increase, and, then, decrease. 
	Hence, network operators can obtain statistical significance regions for chain and star motifs that map to the optimal content dissemination performance. 
	Furthermore, using the obtained regions and the analytical expressions for statistical significance, network operators can effectively identify which clusters of devices can be leveraged for D2D communications while determining the number of serving devices in each identified cluster.}

\end{abstract} 
%

\section{Introduction}
The number of mobile devices, such as smart-phones and tablets, has significantly increased, and almost half a billion mobile devices were added in 2016 \cite{Cisco}. 
This growth in mobile devices has been accompanied by a rise in the need for pervasive wireless connectivity in which users require to access their content of interest anywhere, anytime, and on any device.
One effective approach for such large-scale content dissemination is to use device-to-device (D2D) communication links, which can enable mobile devices to communicate directly without infrastructure \cite{he2017spectral,xiao2014spectrum,liu2015device,mozaffari2016unmanned,zeng2018joint}. 
Compared with broadcasting contents by base stations (BSs), using D2D communications for content dissemination can help better serve users at cell edges and users that are not in the coverage of BSs \cite{xiao2014spectrum,liu2015device,mozaffari2016unmanned,zeng2018joint,bai2016caching,semiari2015context,zhao2017social}.  
Moreover, as each user can have different interests on contents, the information broadcasted by BSs may not be useful for every user \cite{agarwal2015large,wu2015exploiting,meng2017cooperative}. 
In addition, D2D-based content dissemination can offload traffic from BSs and, thus, help alleviate the network congestion \cite{andreev2015analyzing,afshang2016fundamentals,sakr2015cognitive,bacstuǧ2015cache}.
However, 
when deploying D2D communications for content dissemination, one must address two key challenges \cite{xiao2014spectrum,liu2015device,mozaffari2016unmanned,zeng2018joint,bai2016caching,semiari2015context,zhao2017social,agarwal2015large,wu2015exploiting,meng2017cooperative,andreev2015analyzing,afshang2016fundamentals,sakr2015cognitive,bacstuǧ2015cache}: a) designing \textit{content dissemination strategies} for choosing the number of serving devices that can serve as seeds for disseminating contents, and b) developing metrics to assess the content dissemination performance under different network topologies. 


First, in order to choose the optimal set of serving devices, one can
leverage a communication graph composed of D2D links and make use of graph-theoretic properties to identify the most influential devices that can store popular content and to choose the number of serving devices that maximizes the overall system throughput \cite{bai2016caching,semiari2015context,zhao2017social,agarwal2015large,wu2015exploiting,meng2017cooperative}.
For example, in \cite{bai2016caching} and \cite{semiari2015context}, the authors use degree centrality, defined as the number of connected D2D links for every node, and identify the set of nodes with high degree centrality as serving devices. 
Meanwhile, the works in \cite{zhao2017social} and \cite{agarwal2015large} focus on the betweenness centrality, which measures the fraction of shortest paths passing through a focal node, to identify the set of serving devices for D2D content dissemination purposes.
Moreover, the authors in \cite{wu2015exploiting} use the closeness centrality, computed by the sum of distance between a node and all other nodes in the graph, to establish the strategy for choosing the best set of serving devices. 
Furthermore, in \cite{meng2017cooperative}, the authors first build the adjacency matrix for D2D communication network, and then choose the set of serving devices by using the eigenvector corresponding to the largest eigenvalue in the adjacency matrix.

Although the works in \cite{bai2016caching,semiari2015context,zhao2017social,agarwal2015large,wu2015exploiting,meng2017cooperative} exploit the graph-based properties, these works solely focus on the low-order network connectivity, which can only capture the features at the level of individual nodes and edges. 
In other words, these properties just measure the accumulation of influence from one device on other devices or D2D links. 
However, for content dissemination, understanding how information propagates among multiple nodes is critical, as such propagation can directly capture the influence of one device on a group of devices. 
Moreover, the works in \cite{bai2016caching,semiari2015context,zhao2017social,agarwal2015large,wu2015exploiting,meng2017cooperative} do not conduct performance analysis for content dissemination under different network topologies.

To analyze the performance of content dissemination using D2D and derive tractable performance metrics for coverage and data rate, it has become customary to use stochastic geometry techniques \cite{andreev2015analyzing,afshang2016fundamentals,sakr2015cognitive,bacstuǧ2015cache,yang2014green,yang2016heterogeneous}. 
For example, the authors in \cite{andreev2015analyzing} derive the outage probability for an arbitrary D2D link in a model where D2D receivers are uniformly distributed in a circular region around the transmitters.
Moreover, in \cite{afshang2016fundamentals}, to capture the notion of \textit{device clustering}, the authors model the locations of the devices as a Poisson cluster process (PCP), and develop expressions for coverage probability and area spectral efficiency. Furthermore, performance metrics, such as outage probability and average data rate, are also analyzed for systems where the distributions of BSs and devices follow the Poisson point process (PPP) as done in \cite{sakr2015cognitive,bacstuǧ2015cache,yang2014green,yang2016heterogeneous}.


Nevertheless, these works in \cite{andreev2015analyzing,afshang2016fundamentals,sakr2015cognitive,bacstuǧ2015cache,yang2014green,yang2016heterogeneous} do not consider multicast and multi-hop communications, which can extend the range of content dissemination and achieve a better performance than single-hop D2D links \cite{asadi2014survey}. 
In addition, these works are restricted to the modeling and analysis of D2D communication, and do not consider any content dissemination policy or the choice of the optimal set of serving devices.
Although multicast and multihop communications have been considered, respectively, in \cite{lin2014modeling} and \cite{dai2017analytical}, these works do not study more practical scenarios in which both types of communications exist.
Moreover, the authors in \cite{lin2014modeling} and \cite{dai2017analytical} are restricted to D2D systems with uniformly and independently distributed BSs and devices. 
However, due to the similarity of users' content interests, devices
can group together within given areas, e.g., libraries, and form clusters instead of being independently and uniformly distributed \cite{afshang2016fundamentals}.
The main contribution of this paper is to propose a graph-based D2D network analysis framework to determine the content dissemination strategy in clustered D2D networks where both multicast and multi-hop communications exist. 
In particular, we first develop a framework, based on stochastic geometry, that models the distribution of device groups within the D2D network as a cluster process.
Then, we build a D2D communication graph based on the distance distribution among devices and explore frequently occurring network subgraphs, also known as \textit{graph motifs}, to capture content propagation patterns in a group of devices. 
Next, we theoretically analyze the occurrences for two types of graph motifs, i.e., chain (multi-hop) and star (multicast) motifs, and conduct rigorous performance analysis for D2D and cellular devices.
Finally, based on the theoretical analysis, we determine guidelines for designing effective content dissemination strategies in D2D-enabled cellular networks.
To our best knowledge, this is \textit{the first work that exploits network motifs to analyze the performance of content dissemination in D2D-enabled cellular networks}. The novelty of this work lies in the following key contributions:

\begin{itemize}
	\item We explore occurring network structures, known as graph motifs, to capture content propagation patterns among a group of devices and theoretically analyze the occurrences of different motifs in D2D networks. In particular, we model the distribution of devices as a Thomas cluster process (TCP), where the devices are normally scattered around the central points. Then, based on the distance distribution between two arbitrary devices in the network, we formulate a distance-based graph. 
	Also, we derive tractable expressions for the statistical significance to capture the occurrences of two types of motifs, chain and star motifs that respectively capture multi-hop and multicast communications.   
	\item We conduct a comprehensive performance analysis for both D2D chain and star motifs. 
	In particular, we derive closed-form and tractable expressions for the average throughput for D2D and cellular devices, in presence of motifs.   	 
	Moreover, we also derive the outage probabilities for both chain and star motifs.
	\item Extensive simulation results are used to corroborate the analytical results. 
	In particular, we can observe the influence of different system topologies on the occurrence and the outage probability of motifs and the system throughput. Moreover, the results highlight that as the statistical significance of motifs increases, the system throughput will initially increase, and, then, decrease.
	\item The proposed framework provides important guidelines for designing effective content dissemination strategies in D2D-enabled cellular networks. 
	In particular, using the derived expressions and the simulation results, network operators can determine the statistical significance regions for different motifs that map to the optimal system throughput. 
	Moreover, by comparing the motifs' statistical significance in each cluster with the optimal regions, the operators can identify which clusters can be optimally leveraged for D2D communications. 
	In addition, based on the analytical derivations on the statistical motif occurrence, one can also determine the number of serving devices in each identified cluster. 
\end{itemize}

The rest of the paper is organized as follows. Section \uppercase\expandafter{\romannumeral2} presents the system model and key assumptions. In Section \uppercase\expandafter{\romannumeral3} and Section \uppercase\expandafter{\romannumeral4}, we conduct graph motifs based analysis for D2D networks and develop the performance-related metrics. Simulation results are provided in Section \uppercase\expandafter{\romannumeral5} and conclusions are drawn in Section \uppercase\expandafter{\romannumeral6}.
\label{sec:intro}

\section{System Model}
\begin{figure}[!t]
	\centering
	\subfloat[Spatial motifs in the representative cluster.]{\includegraphics[width=0.48\textwidth]{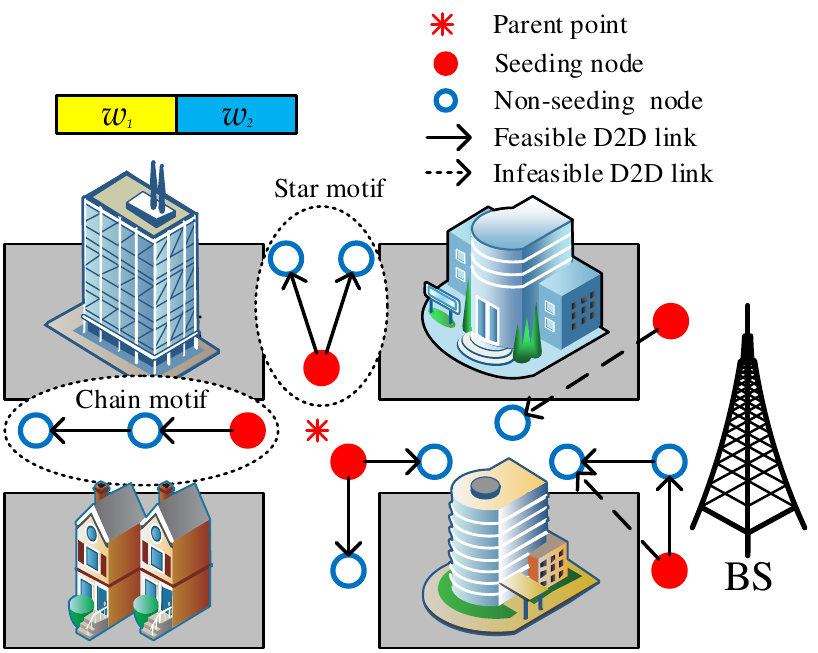}\label{fig:f2}}\hfill
	\subfloat[Different clusters existing within a geographical area.]{\includegraphics[width=0.48\textwidth]{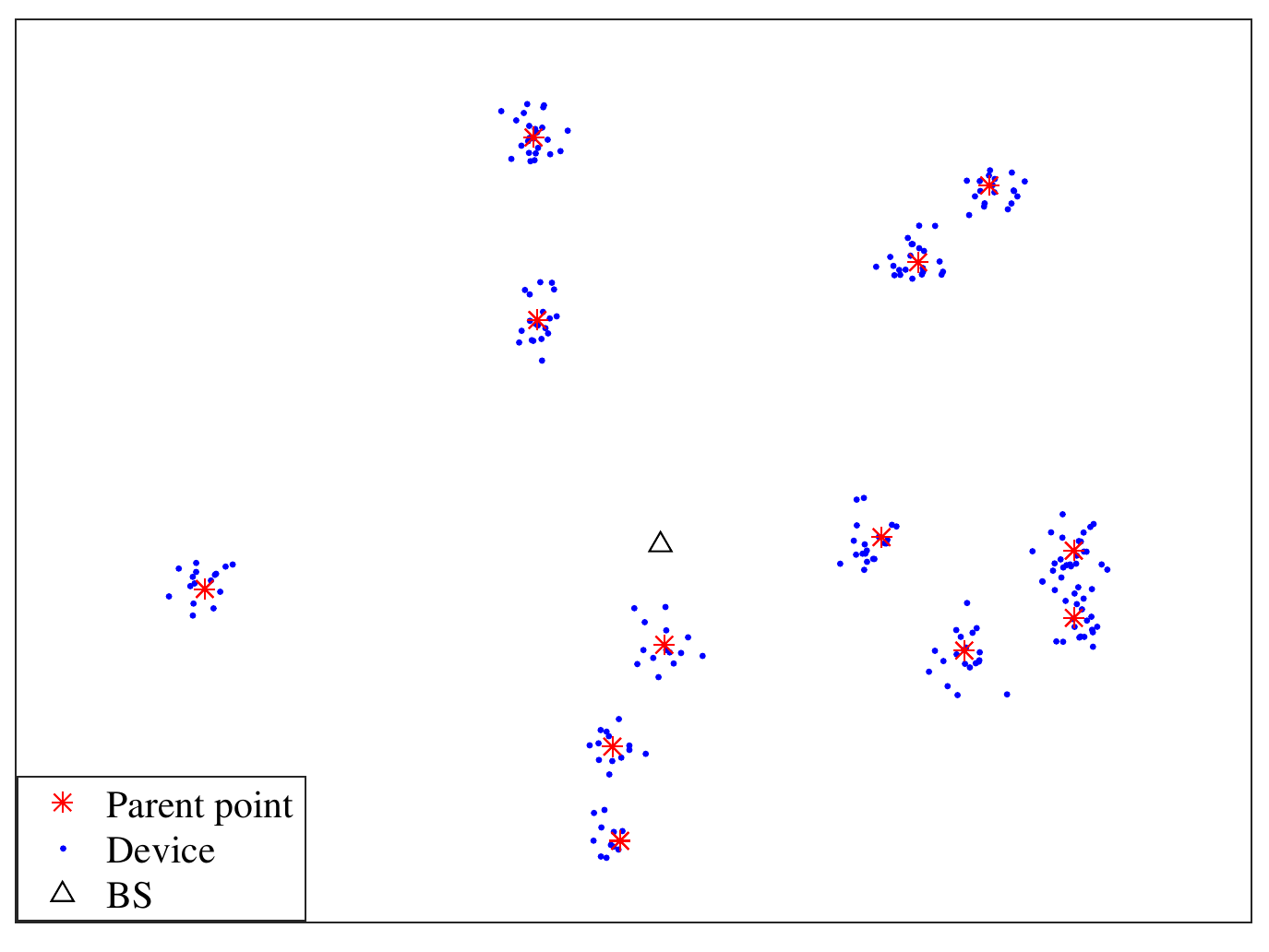}\label{fig:f1}}
	\hfill
	\caption{System model that follows a TCP.}
	\label{fig:1f1}
	\vspace{-0.25in}
\end{figure}

Consider a cellular network composed of a single BS and a number of devices that can communicate with the BS over cellular links as well as directly exchange information with one another via D2D communications, as shown in Fig. \ref{fig:1f1}\subref{fig:f2}. 
Here, devices with injected contents from the BS are referred to as  \textit{seeding nodes}, and those that obtain the content from nearby devices are called \textit{non-seeding nodes}. 
Moreover, due to the similarity of users' interests, devices can 
concentrate at the same area, such as a library or a stadium, and the grouped devices will form clusters \cite{semiari2015context,zhao2017social,agarwal2015large,wu2015exploiting,meng2017cooperative}. 
Such a cluster-based model has been recently introduced in \cite{wu2015exploiting} as an effective and practical approach to model D2D communication networks and study content dissemination.  
In addition, as done in \cite{zhou2013intracluster}, to avoid the interference between cellular and D2D links, we consider a D2D overlay system in which the available bandwidth $w$ is partitioned into $w_{1}\!\!=\!\!\beta w$ for D2D links and $w_{2}=(1\!\!-\!\!\beta)w$ for cellular links, where $\beta \in (0,1)$.

\label{section_system_model}
To capture the non-uniform, cluster-based distribution of devices in the network, we consider a TCP composed of a parent point process and a daughter point process \cite{haenggi2012stochastic}. 
In particular, as shown in Fig. \ref{fig:1f1}\subref{fig:f1}, we model the distribution of the parent points (center points of the clusters) as a PPP $\Phi_{p}$ with density $\lambda_{p}$. 
Also, the daughter points (devices) in each cluster are normally scattered with variance $\sigma^{2}$ around the corresponding parent point $x \in \Phi_{p}$  \cite{haenggi2012stochastic}. 
Note that the locations of daughter points, with respect to the corresponding parent point, are independent and identically distributed (i.i.d.) variables \cite{haenggi2012stochastic}.
\vspace{-0.5cm}

\subsection{D2D Communication Graph}
\vspace{-0.2cm}
\label{subsectionD2D}
As shown in Fig. \ref{fig:1f1}\subref{fig:f2}, we arbitrarily select a cluster with parent point $x_{r} \in \Phi_{p}$, considered as \textit{a representative cluster $C_{r}$}. 
In cluster $C_{r}$, seeding nodes can disseminate popular contents to other devices via D2D links. 
Nonetheless, due to limited transmit power and different interests on contents, devices belonging to cluster $C_{r}$ will not communicate with other clusters.
Hence, in cluster $C_{r}$, we can model the D2D network as a directed acyclic graph $G\!\!=\!\!(\Phi_{x_{r}}, \mathcal{E}_{x_{r}})$, whose vertices are devices in the set $\Phi_{x_{r}}$ and whose edges, within the set $\mathcal{E}_{x_{r}}$, are D2D links among devices in the cluster. 
In particular, each D2D link can be defined as a $2-tuple$ edge, $e=<\!i,j\!>$, $i,j \in \Phi_{x_{r}}$, where device $i$ and device $j$ are the transmitter and the receiver, respectively. 
Here, as the existence of a D2D link between a pair of devices is determined by many factors, such as social or business relationship \cite{semiari2015context} and distance \cite{frlan2000direct1}, we assume that a device may communicate with any arbitrary nearby device as long as the distance between them is below the maximum allowable communication distance $s_{\textrm{th}}$ as done in \cite{frlan2000direct1} and \cite{ding2014feasible}.
Moreover, if there exists a graph $G'\!\!=\!\!(\Phi_{x_{r}}', \mathcal{E}_{x_{r}}')$ with $\Phi_{x_{r}}'\!\!\subseteq\!\!\Phi_{x_{r}}$ and $\mathcal{E}_{x_{r}}'\!\!\subseteq\!\!\mathcal{E}_{x_{r}}$, the graph $G'$ is a subgraph of $G$. 
For analytical tractability, we assume that the number of devices in each cluster is $N$. 
As will be clear from the following discussion, although the total number of devices in each cluster is the same, the number of active D2D links varies from one cluster to another, which maintains the generality of the model.

By focusing on cluster $C_{r}$, we can observe that groups of D2D links can form various subgraphs with different structures, which can show how the content propagates among a group of devices. 
For example, as shown in Fig. \ref{fig:1f1}\subref{fig:f2}, two D2D links can share the same transmitter, and the receiver of one link can act as the transmitter of another D2D link. 
To characterize D2D subgraphs appearing in the network, we exploit the notion of \textit{graph motifs}, defined as subgraphs that appear in a network significantly more than in a baseline system, such as the random graph where edges are randomly assigned to a pair of arbitrary devices \cite{milo2002network}.
In particular, as a customary for the study of motifs, we use the \emph{Z-score} to measure the statistical significance of different motifs in the clustered-based network and the $Z$-score is defined as \cite{alon2007network}:
 
	\begin{align}
\label{f8}
Z=\frac{c_{\textrm{o}}-c_{\textrm{r}}}{\varepsilon_{\textrm{r}}},
\end{align}
where $c_{\textrm{o}}$ is the number of occurrences of a given motif in the network, and $c_{\textrm{r}}$ and $\varepsilon_{\textrm{r}}$ denote, respectively, the mean and the standard deviation of motif's occurrence in the baseline system. 
For the baseline system, we use a random graph model that shares the same total number of D2D links existing in graph $G$, but randomly rewires each link between an arbitrary pair of devices \cite{kolascyk2013statistical}.
Note that, a positive $Z$-score means that the subgraph is not likely to randomly occur without following any pattern. 
Meanwhile, a larger $Z$-score implies a more frequent occurrence of the corresponding motif in the D2D communication network.

In addition, we assume that each seeding device can disseminate content to  multiple non-seeding devices, and non-seeding devices are capable of forwarding the content to others. 
Therefore, as illustrated in Fig. \ref{fig:1f1}\subref{fig:f2}, for any group of three devices, we can observe two types of communication motifs -- the \emph{star motif} and the \emph{chain motif}. 
The star motif can be viewed as a \emph{multicast} link in which the seeding device disseminates its content to two non-seeding devices, and is captured by two edges sharing the same first element in the communication graph.
Meanwhile, the chain motif can be considered as a two-hop D2D link and we also assume that the chain motif follows a decode-and-forward relaying protocol \cite{hasna2003optimal}. 
In particular, the seeding device will first send its content to a non-seeding device in one time slot, which functions as a relay propagating the content to another non-seeding device in the next time slot. 
Similarly, for the chain motif in the graph, there exists two edges connecting three devices, and the second element of one edge is the first element of the next edge. 
Here, we focus on three-device motifs, and, thus, the maximum number of motifs in a cluster is $N_{m}=\lfloor\frac{N}{3}\rfloor$, where any arbitrary two motifs do not share common devices and devices in each group are randomly and uniformly selected.\vspace{-0.6cm}

\subsection{Channel Model and Interference Analysis}
\vspace{-0.2cm}
We model the channel for the D2D link $<i,j>$, $i,j\in \Phi_{x_{r}}$, in cluster $C_{r}$ centred at $x_{r} \in \Phi_{p}$ as a Rayleigh fading channel.
Hence, the received power is given by 
$P_{ij}=P_{t}h_{ij}d_{ij}^{-\alpha}$, 
where $P_{t}$ is the transmit power of each device, $h_{ij}$ is the channel gain that follows an exponential distribution with mean equal to $1$, $d_{ij}$ is the corresponding distance between any two devices $i$ and $j$, and $\alpha$ is the path loss exponent. 
Similar to the bandwidth allocation in \cite{zhang2017network}, in a star motif, we assume that the two D2D links will split the frequency resources evenly and each link uses half of the bandwidth, $\frac{w_{1}}{2}$. Meanwhile, in a chain motif, the first and second D2D links will reuse the same resource, $w_{1}$, at two consecutive time slots. 
Therefore, due to the co-channel deployment of D2D links, the receiving nodes will encounter interference from other D2D links, irrespective of their motif types.
We designate the interference on device $j$ generated from devices in cluster $C_{r}$ as \textit{the intra-cluster interference} $I^{x_{r}}_{j-\textrm{intra}}$, and the interference from devices within other clusters as \textit{the inter-cluster interference} $I^{x_{r}}_{j-\textrm{inter}}$. The two types of interference are given by:
\begin{align}
\label{ff11}
I^{x_{r}}_{j-\textrm{intra}} = \sum_{y\in \Phi_{x_{r}} \setminus \{i,j\}}\mathbbm{1}_{y}P_{t}h_{yj}d_{yj}^{-\alpha}, \hspace{0.1cm}
I^{x_{r}}_{j-\textrm{inter}} = \sum_{x\in \Phi_{p}\setminus x_{r}} \sum_{y \in \Phi_{x}}\mathbbm{1}_{y}
P_{t}h_{yj}d_{yj}^{-\alpha},
\end{align}
where the indicator variable $\mathbbm{1}_{y}=1$ if device $y$ is transmitting data to other devices via D2D link; otherwise, $\mathbbm{1}_{y}=0$.
For a large D2D network, the thermal noise power at the receiver will be substantially smaller than the received interference power \cite{afshang2016modeling}. Thus, the expression of the signal-to-interference-plus-noise ratio  (SINR) of the D2D link can be approximated to the signal-to-interference ratio (SIR) as follows
\begin{align}
\label{ff1}
\delta_{ij} \approx \frac{P_{t}h_{ij}d_{ij}^{-\alpha}}{I^{x_{r}}_{j-\textrm{intra}}+I^{x_{r}}_{j-\textrm{inter}}}.
\end{align}
Since we consider a single-cell scenario and the BS will transmit contents to seeding devices in different time slots, the seeding devices will experience no interference and their signal-to-noise ratio (SNR) will be:
\begin{align}
\label{ff2}
\delta_{bj} = \frac{P_{b}h_{bj}d_{bj}^{-\alpha}}{P_{n}},
\end{align}
where $P_{b}$ is the transmission power of the BS, $h_{bj}$ is the channel gain that follows an exponential distribution with mean equal to $1$, $d_{bj}$ represents the distance between the BS and device $j$, and $P_{n}$ captures the power of the background noise. 
Moreover, we can find the achievable data rate $R$ for D2D links and cellular links, according to $R\!\!=\!\!w_{3}\log_{2}(1\!\!+\!\!\delta)$, where $w_{3}$ is the assigned bandwidth. 
Note that, we can also extend our work to a multiple-cell model.
To do so, one must take into account the additional interference generated by the different BSs and devices under the coverage of these BSs.
\vspace{-0.6cm}
\subsection{Relationship between the Occurrence of Motifs and the D2D System Throughput}
\vspace{-0.2cm}
Since the number of D2D links in each three-node motif is two, the expected
number of D2D links in the baseline system is $2c_{\textrm{o}}$. 
The probability of one pair of arbitrary devices being connected by a D2D link in the baseline system is thereby $\mathbb P_{r}=\frac{2c_{\textrm{o}}}{{{N_{m}}\choose{2}}}$. 
In this case, the probability with which three devices in the baseline system form either a chain motif or a star motif is $\mathbb P_{m}={{3}\choose{1}}\mathbb P_{r}^2(1-\mathbb P_{r})$. 
Moreover, we can verify the mean $c_{r}=N_{m} \mathbb P_{m}$ and the standard deviation $\sigma_{r}=\sqrt{N_{m}\mathbb P_{m}(1-\mathbb P_{m})}$ of a motif's occurrence in the baseline system.
Furthermore, based on (\ref{f8}), we can also derive $\frac{dZ}{dc_{o}}>0$, when $c_{o}\geq 1$, indicating that the $Z$-score is an increasing function of the occurrence $c_{o}$ of the motif.
 
Due to the co-channel deployment of D2D links, interference will increase as the frequency of occurrence of any given motif increases. 
Also, as the $Z$-score is an increasing function of the occurrence of a motif, interference will increase when the $Z$-score increases. 
Thus, for clusters having higher $Z$-scores, content dissemination via D2D communications will experience a high interference, thereby degrading the dissemination throughput. 
On the other hand, for clusters with a smaller $Z$-score, only a limited number of D2D links can be formed, resulting in a low spectral reuse. 
Therefore, we need to find the $Z$-score region, which can yield an optimal system performance. 
Given such a $Z$-score region, the network operator can identify clusters that can be used for D2D communications by just comparing any given $Z$-score of the clusters with the optimal region. 
Moreover, as the $Z$-score is directly linked to the occurrence of motifs in the communication network, the number of seeding devices in each identified cluster can be also determined.

However, due to the lack of closed-form expressions that link the $Z$-score and the wireless throughput, directly finding the $Z$-score region that yields the optimal system performance is challenging.
Nevertheless, we can observe that the value of the $Z$-score depends on number of devices $N$, their locations, and the maximum communication distance $s_{\textrm{th}}$.
In addition, the system throughput is also a function of these three parameters.
Therefore, instead of finding a direct relationship between the $Z$-score and the system throughput, one can alternatively determine the analytical expressions of the $Z$-score and the system throughput as functions of their common parameters. 
Using such analytical expressions, the BS can calculate the $Z$-score and the system throughput corresponding to the same parameter settings,  
and further observe how the system throughput changes as the $Z$-score of motifs varies. 
The changes of the system throughput as function of the $Z$-score can enable network operators to disclose the relationship between the occurrence of motifs and the system performance, and further determine the $Z$-score region which maps to the optimal system throughput. 
Next, we will first leverage the distance distribution among devices and the stochastic properties inherent in the TCP to derive tractable expressions for the $Z$-scores of chain and star motifs and the system throughput. 
Also, we will further identify the hidden relationship between motifs and the system performance via numerical results and determine the content dissemination strategy for clustered networks. 

\vspace{-0.5cm}
\section{Statistical Significances of Star and Chain Motifs }
\vspace{-0.2cm}
\label{$Z$-score}
In this section, for an arbitrarily chosen group of three devices in cluster $C_{r}$, we use the distance distribution between any two randomly and uniformly selected points to derive the probability of the content dissemination pattern among three devices being either a star or chain motif. 
Based upon the probability of occurrence, we can derive closed-form expressions for the $Z$-scores.
\vspace{-0.6cm}
\subsection{Distance Distribution}
\vspace{-0.2cm}
\label{subsect1}

We arbitrarily select a group of three devices in cluster $C_{r}$, and due to the stationarity of the TCP, we treat an arbitrary device out of these three devices as \textit{a typical point} located at the origin. 
In addition, we assume that the location of the parent point is at $x_{r}\in \Phi_{p}$, and the other two devices are located at $y_{1}$ and $y_{2}$ relative to the parent point. 
Therefore, according to the definition of the TCP in Section  \ref{section_system_model}, $y_{1}$ and $y_{2}$ are distributed according to a symmetric normal distribution with the variance $\sigma^{2}$ around the parent point $x_{r}$, and the parent point also follows a zero mean symmetric normal distribution with the variance $\sigma^{2}$ relative to the origin point. 
Additionally, $x_{r}$, $y_{1}$, and $y_{2}$ are i.i.d. variables, $x_{r}, y_{1}, y_{2} \in \R^{2}$.

Furthermore, we denote $\{\mathcal{D}_{i}\}_{i=1,\ldots,N-1}$ as the distance set from the typical point to another uniformly selected point in the same cluster.
In particular, the realization of $\{\mathcal{D}_{i}\}_{i=1:N-1}$ is $d_{i}=||x_{r}\!+\!y_{i}||, i\!=\!1,\ldots,N-1$. 
Conditioned on the distance from the typical point to the parent point $s_{r}=||x_{r}||$, the probability density function (PDF) of $d_{i}$ follows a Rician distribution \cite{afshang2016modeling}: 
\begin{align}
\label{f2}
f_{D}(d_{i},s_{r};\sigma^{2})=\frac{d_{i}}{\sigma^{2}}\exp\bigg(-\frac{d_{i}^{2}+s_{r}^{2}}{2\sigma^{2}}\bigg)I_{0}\bigg(\frac{d_{i}s_{r}}{\sigma^{2}}\bigg), \space d_{i}, s_{r}\geq0,
\end{align}
where $I_{0}(\cdot)$ is the modified Bessel function of first kind with zero order. 
Conditioned on $s_{r}=||x_{r}||$, the elements in $\{\mathcal{D}_{i}\}$ are i.i.d. variables as shown in \cite{afshang2016modeling}. 
Moreover, the authors in \cite{afshang2016modeling} also conclude that, when the distances from the typical point to the parent points of other clusters are a fixed value, the distances between the typical device at the origin and the devices in other clusters are also i.i.d. variables and follow Rician distributions. As it will be clear from the following discussion, we use this property to calculate inter-cluster interference.
In addition, according to the definition of the TCP, the distance $s_{r}$ follows a Rayleigh distribution with scale parameter $\sigma$ and with the following PDF: 
\begin{align}
\label{f3}
f_{S}(s_{r};\sigma^{2})=\frac{s_{r}}{\sigma^{2}}\exp\bigg(-\frac{s_{r}^{2}}
{2\sigma^{2}}\bigg), \space s_{r}\geq0.
\end{align}

\vspace{-0.6cm}
\subsection{Probability of Occurrence of Motifs}
\vspace{-0.2cm}
Based on the distance distribution obtained in Section \ref{subsect1}, we can derive the probability with which a group of three arbitrary devices will form either a chain or star motif. 
First, we can observe that, for both chain and star motifs, there is a node capable of directly communicating with the other two devices. 
Hence, if we consider a device with two direct D2D links as a typical point located at the origin, we can express the distance from the typical point to the other two uniformly and randomly selected devices as the norm of $m=x_{r}+y_{1} \in \mathbb{R}^{2}$, and $n=x_{r}+y_{2}\in \mathbb{R}^{2}$, where $m=(m^{(1)},m^{(2)})$, $n=(n^{(1)},n^{(2)})$, and $x_{r}=(x_{r}^{(1)},x_{r}^{(2)})$. 
According to the definition of a TCP, $x_{r}+y_{1}$ and $x_{r}+y_{2}$ are zero mean complex Gaussian random variables with variance $2\sigma^{2}$. 
Thus, $s_{1}=||m||$ and $s_{2}=||n||$ follow the Rayleigh distribution with scale parameter $\sqrt{2}\sigma$. 
Due to the common element, $x_{r}$, in $m$ and $n$, the distances $s_{1}$ and $s_{2}$ will be correlated. 
Based on the distribution and correlation between $s_{1}$ and $s_{2}$, next, we can obtain the joint probability of both $s_{1}$ and $s_{2}$ being smaller than $s_{\textrm{th}}$.

\begin{theorem} 
	\label{lemma1}
	Given that $s_{1}\!=\!||m||$ and $s_{2}\!=\!||n||$ follow Rayleigh distributions and are correlated, the joint probability of both $s_{1}\!=\!||m||$ and $s_{2}\!=\!||n||$ being smaller than $s_{\textrm{th}}$ is:
	\begin{align}
	\label{f7}
	\normalfont \mathbb P
	_{S,S}(s_{\textrm{th}},s_{\textrm{th}})
	=\frac{3}{4}\sum_{k=0}^{\infty}\left( \left(\frac{1}{2}\right)^{k}\frac{\gamma\left(1+k,\frac{s_{\text{th}}^{2}}{3\sigma^{2}}\right)}{(k!)} \right)^{2}, 
	\end{align}
where $\gamma(\cdot,\cdot)$ refers to the lower incomplete gamma function, defined as $\gamma(a,b)=\int_{0}^{b}c^{a-1}e^{-c}dc$.
\begin{proof}[Proof:\nopunct]
	See Appendix \ref{appendixtotheorem1}.
\end{proof}
\end{theorem}

\begin{figure}[!t]
	\centering
	\includegraphics[width=3in,height=0.8in]{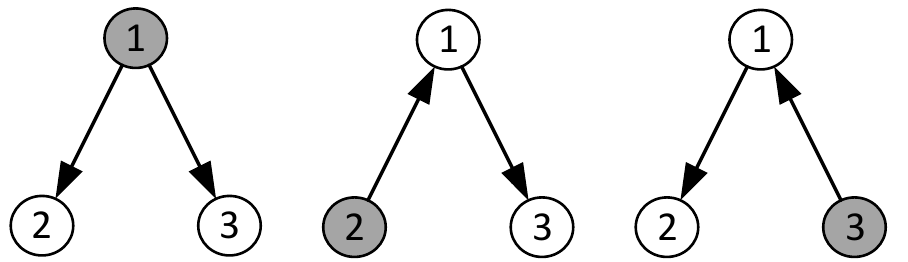}
	\DeclareGraphicsExtensions.
	\vspace{-0.1in}
	\caption{Possible scenarios of three-node motifs. Node 1 is a typical node capable of directly communication with nodes 2 and 3. Nodes shown in grey color are seeding devices.}
	\label{possibleThreeNodeMotif}
	\vspace{-0.3in}
\end{figure}
In (\ref{f7}), $\mathbb P_{S,S}(s_{\textrm{th}},s_{\textrm{th}})$ represents the probability that a typical device can build D2D links with two other devices uniformly and randomly chosen from the cluster, irrespective of the link direction. 
Since a receiving device cannot access the content from multiple transmitters, and a transmitter can spread content to at most two devices simultaneously, $\mathbb P_{S,S}(s_{\textrm{th}},s_{\textrm{th}})$ is the probability with which any three devices can form either the star motif or the chain motif. 
For ease of exposition, we use the term ``three-node motifs'' to refer to the union of the star and chain motifs, and replace $\mathbb P_{S,S}(s_{\textrm{th}},s_{\textrm{th}})$ with $\mathbb P_{S,S}$ hereinafter. 
Moreover, as shown in Fig. \ref{possibleThreeNodeMotif}, there are three possible scenarios for a random three-node motif where two possible scenarios are chain motifs and one scenario is the star motif. 
Without loss of generality, we assume that each node in the three-node motif can be the seeding node with equal probability.
In other words, each scenario will occur with equal possibility.
In this case,  the probability of one three-node motif being a star motif is $\theta=\frac{1}{3}$, and the probability of forming a chain motif will be $1\!-\!\theta=\frac{2}{3}$. 
\vspace{-0.6cm}
\subsection{$Z$-scores for chain and star motifs}
\vspace{-0.2cm}
To calculate the $Z$-scores for the chain and star motifs, we must derive the occurrence of these two motifs in cluster $C_{r}$ and compare it with the baseline system.
First of all, we can express the probability of $n$ groups of three-node motifs existing in a cluster as 
${{N_{m}}\choose{n}}(\mathbb P_{S,S})^{n}(1-\mathbb P_{S,S})^{N_{m}-n}, n\leq N_{m}$.
Using the properties of the binomial distribution, the expected number $c_{\textrm{o}}$ of three-node motifs in the network with distance will be 
$c_{\textrm{o}}=N_{m} \times  \mathbb P_{S,S}$.
Accordingly, the expected numbers of occurrence for the star and chain motifs are given by
\begin{align}
\label{occcccc}
c_{\textrm{o}}^{\textrm{star}}=\theta N_{m} \times \mathbb P_{S,S}, \hspace{0.1cm}
c_{\textrm{o}}^{\textrm{chain}}=(1-\theta)N_{m} \times  \mathbb P_{S,S}.
\end{align}%

For the baseline system, we can derive the probability of one pair of arbitrary devices being connected by a D2D link as follows:
\begin{align}
\label{f11}
\mathbb P_{r}=\frac{2c_{\textrm{o}}}{{{N_{m}}\choose{2}}}=\frac{4\mathbb P_{S,S}}{N_{m}-1}.
\end{align}%
In this case, the probability with which three devices form a three-node motif is ${{3}\choose{1}}(\mathbb P_{r})^{2}(1-\mathbb P_{r})$. Furthermore, the probabilities of being a star motif or a chain motif in the baseline system are $3\theta(\mathbb P_{r})^{2}(1-\mathbb P_{r})$, and $3(1-\theta)(\mathbb P_{r})^{2}(1-\mathbb P_{r})$, respectively. 
Similarly, by using the binomial distribution, we can express the mean $c^{\textrm{star}}_{\textrm{r}}$ and the standard deviation $\varepsilon^{\textrm{star}}_{\textrm{r}}$ for the star motif and the counterparts $c^{\textrm{chain}}_{\textrm{r}}$ and $\varepsilon^{\textrm{chain}}_{\textrm{r}}$ for the chain motif in the baseline system as 
\begin{align}
\label{f12}
&c^{\textrm{star}}_{\textrm{r}}\!=\!\frac{48\theta N_{m}(\mathbb P_{S,S})^{2}(N_{m}-1-4\mathbb P_{S,S})}{(N_{m}-1)^{3}}, \\ \label{f1221}
&c^{\textrm{chain}}_{\textrm{r}}\!=\!\frac{48(1-\theta)N_{m}(\mathbb P_{S,S})^{2}(N_{m}-1-4\mathbb P_{S,S})}{(N_{m}-1)^{3}}, \\ \label{f122}
&\varepsilon^{\textrm{star}}_{\textrm{r}}\!=\!\sqrt{\frac{48\theta N_{m}(   \mathbb P_{S,S})^{2}(N_{m}-1-4\mathbb P_{S,S})}{(N_{m}-1)^{3}}\left(1-\frac{48\theta N_{m}( \mathbb P_{S,S})^{2}(N_{m}-1-4\mathbb P_{S,S})}{(N_{m}-1)^{3}}\right)}, \\ \label{f123}
&\varepsilon^{\textrm{chain}}_{\textrm{r}}\!=\!\sqrt{\frac{48(1\!-\!\theta)N_{m}(  \mathbb P_{S,S})^{2}(N_{m}\!-\!1\!-\!4\mathbb P_{S,S})}{(N_{m}-1)^{3}}\left(1\!-\!\frac{48(1\!-\!\theta)N_{m}( \mathbb P_{S,S})^{2}(N_{m}\!-\!1\!-\!4\mathbb P_{S,S})}{(N_{m}-1)^{3}}\right)}.
\end{align}%

After obtaining the occurrence information for both motifs in cluster $C_{r}$ and the baseline system, we can derive the $Z$-scores for both motifs. 
\begin{remark}
	\label{lemma3}
	By using the analytical results in (\ref{f8}), (\ref{occcccc}) and (\ref{f12})-(\ref{f123}), the expressions of $Z$-scores for both motifs in the D2D system can be determined. In fact, (\ref{occcccc}) captures the expected number of occurrences for chain and star motifs in cluster $C_{r}$. 
	Also, (\ref{f12})-(\ref{f123}) represent the mean and standard variance of the occurrences for both motifs in the baseline system.
\end{remark}

As observed from Theorem \ref{lemma1}, the joint probability $\mathbb P_{S,S}$ is a function of the total number of devices $N$, the distribution variance $\sigma^{2}$, and the maximum allowable distance $s_{\textrm{th}}$. Therefore, we can conclude that the value of the $Z$-score is also dependent on $N$, $\sigma^{2}$, and $s_{\textrm{th}}$. 
Note that the network operator is capable of estimating parameters, such as the variance $\sigma^{2}$, for the distribution of devices in each cluster based on the long-term observation \cite{baddeley2008analysing}.
In this case, the operator can calculate the $Z$-score for each motif without collecting data from devices, decreasing the uplink traffic overhead especially when the number of devices is large.

\section{Expected Average Throughput and Outage Probability}
\label{throughput}
To analyze the performance of the content dissemination strategy under different clustered network topology parameters, we use the distance distribution properties inherent in the TCP to calculate the expected throughput for D2D and cellular devices. Moreover, the outage probabilities for the chain and star motifs are also derived. 

\begin{figure}[!t]
	\centering
	\includegraphics[width=2.8in,height=2.1in]{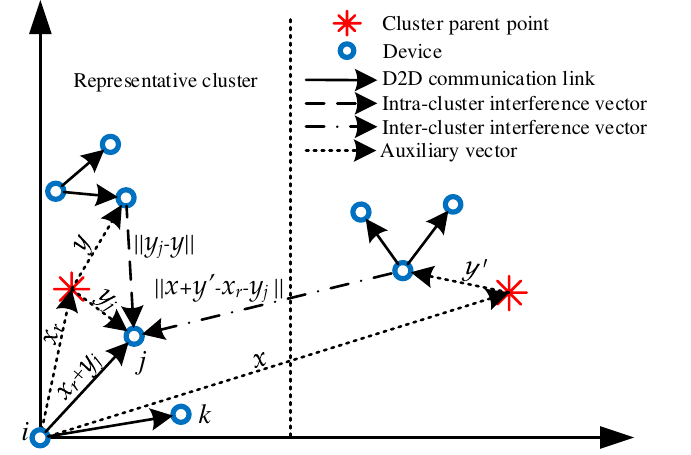}
	\DeclareGraphicsExtensions.
	\vspace{-0.1in}
	\caption{Illustration of the inter- and intra-cluster interference to the representative star motif.}
	\label{Illustrate2}
	\vspace{-0.3in}
\end{figure}
\vspace{-0.6cm}
\subsection{Expected Throughput for Non-Seeding Devices in the Star Motif}
\vspace{-0.2cm}
As shown in Fig. \ref{Illustrate2}, we choose an arbitrary star motif as \textit{a representative star motif $M_{s}$} in cluster $C_{r}$. 
In motif $M_{s}$, the two receiving devices, $j$ and $k$, will access data via D2D links from the seeding device, $i$, located at the origin. 
Moreover, we assume that the set of seeding devices in star motifs as $\mathcal{T}_{x}$ in the cluster $x \in \Phi_{p}$, and the number of elements in $\mathcal{T}_{x}$ cannot be greater than $N_{m}$. 

Next, we take device $j$ as an example and conduct Laplace transforms for the intra- and inter-cluster interference encountered by device $j$. 
Based on the derived Laplace transforms, we can study the performance-related metrics, such as the outage probability and the expected throughput. 
According to the stationarity of the TCP and the symmetry between the two receiving devices in the star motif, the Laplace transforms of the intra- and the inter-cluster interference for device $j$ can apply to another device $k$ and receiving devices in other star motifs. 

\begin{lemma}
	\label{lemmafirst}
	For device $j$ in the star motif, the Laplace transform of the intra-cluster interference can be expressed as  
	\begin{align}
	\label{intra-star}
	\mathcal{L}^{\emph{\textrm{star}}}_{j\emph{\textrm{-intra}}}(s)=&\left(\int_{0}^{\infty}\frac{1}{1+sP_{t}v_{1}^{-\alpha}}f_{V_{1}}(v_{1})dv_{1}\mathbb P_{S,S}+1-\mathbb P_{S,S}\right)^{N_{m}-1},
	\end{align}%
	where $f_{V_{1}}(v_{1})=f_{S}(v_{1};2\sigma^{2})$, $v_{1}\geq 0$.
	\begin{proof}[Proof:\nopunct]
		See Appendix \ref{Appendix B}.
	\end{proof}
\end{lemma}




The Laplace transform of the inter-cluster interference is derived next. 
\begin{lemma}
	\label{lemmasecond}
	For device $j$ in the star motif, the Laplace transform of the inter-cluster interference is given by 
	\begin{align} 
	\label{inter-star}
	\mathcal{L}^{\emph{\textrm{star}}}_{j\emph{\textrm{-inter}}}(s)\!\! =\!\! \exp\left(-\lambda_{p} 2\pi \int_{0}^{\infty}\left(1-\left(\int_{0}^{\infty}\frac{1}{1+sP_{t}v_{2}^{-\alpha}}f_{V_{2}}(v_{2}|t)dv_{2}\mathbb P_{S,S}+1-\mathbb P_{S,S}\right)^{N_{m}}\right)tdt\right),
	\end{align}%
	where $f_{V_{2}}(v_{2}|t)=f_{D}(v_{2},t;3\sigma^{2})$, $v_{2}\geq 0$.
	\begin{proof}[Proof:\nopunct]
		See Appendix \ref{Appendix C}
	\end{proof}
\end{lemma}


Considering that content dissemination within a star motif will not occur an outage if and only if the SIR for both D2D links in the star motif exceed the minimum requirement $\delta_{\textrm{th}}$ \cite{hasna2003optimal}, then the outage probability for the star motif can be derived as follows.

\begin{theorem}
	\label{lemmastar}
	The outage probability for the star motif is given by:
	\begin{align}
	\label{lemma41star}
	\mathbb P^{\emph{\textrm{star}}}_{\emph{\textrm{outage}}}(\delta_{\emph{\textrm{th}}})&= 1-\left( \int_{0}^{\infty}\mathcal{L}^{\emph{\textrm{star}}}_{j\emph{\textrm{-inter}}}\left(\frac{\delta_{\emph{\textrm{th}}}}{P_{t}}r_{1}^{\alpha}\right)\mathcal{L}^{\emph{\textrm{star}}}_{j\emph{\textrm{-intra}}}\left(\frac{\delta_{\emph{\textrm{th}}}}{P_{t}}r_{1}^{\alpha}\right)f_{R_{1}}(r_{1})dr_{1}\right)^{2},
	\end{align}%
	where $f_{R_{1}}(r_{1})=f_{S}(r_{1};2\sigma^{2})$, $r_{1}\geq 0$,
	\begin{proof}[Proof:\nopunct]
	We first calculate the SIR distribution for a D2D link to obtain the probability of a D2D link meeting the minimum SIR requirement: 
		\begin{align}
		\label{f3201star}
		\mathbb P^{\textrm{star}}_{j}(\delta_{\textrm{th}})&=\E(\mathbb{P}\{\delta_{j}>\delta_{\textrm{th}}\})\nonumber \\
		& =\E\left[\mathbb{P}\left(h_{ij}>\frac{\delta_{\textrm{th}}}{P_{t}}(I^{x_{r}}_{j\textrm{-intra}}+I^{x_{r}}_{j\textrm{-inter}})||x_{r}+y_{j}||^{\alpha}\right)\right] \nonumber \\
		&\stackrel{(a)}{=}\E\left[\E\left[\exp\left(-\frac{\delta_{\textrm{th}}}{P_{t}}(I^{x_{r}}_{j\textrm{-intra}}+I^{x_{r}}_{j\textrm{-inter}})||x_{r}+y_{j}||^{\alpha}\right)\right]\right]  \nonumber\\
		&\stackrel{(b)}{=} \int_{0}^{\infty}\mathcal{L}^{\textrm{star}}_{j\textrm{-inter}}\left(\frac{\delta_{\textrm{th}}}{P_{t}}r_{1}^{\alpha}\right)\mathcal{L}^{\textrm{star}}_{j\textrm{-intra}}\left(\frac{\delta_{\textrm{th}}}{P_{t}}r_{1}^{\alpha}\right)f_{R_{1}}(r_{1})dr_{1},
		\end{align}%
		where $(a)$ follows the fact that the channel gain $h_{ij}\!\!\sim\!\!\exp(1)$ for a Rayleigh fading channel. 
		In $(b)$, we substitute $r_{1}$ with $||x_{r}\!\!+\!\!y_{j}||$ and use polar coordinates. 
		Note that the probability in (\ref{f3201star}) can also apply to the link between devices $i$ to $k$, due to the symmetry between devices $j$ and $k$.
		As outage will not occur when the SIR of both links exceeds the threshold $\delta_{\textrm{th}}$, we use the probability in (\ref{f3201star}) for both D2D links in the star motif to derive the outage probability in (\ref{lemma41star}).
	\end{proof} 
\end{theorem}

Since $R_{j}=\frac{w_{1}}{2}\log_{2}(1+\delta_{j})$, $R_{j}<R$ will yield $\delta_{j}<2^{\frac{2R}{w_{1}}}-1$. Thus, after replacing $\delta_{\textrm{th}}$ with $2^{\frac{2R}{w_{1}}}-1$ in (\ref{f3201star}), we can have the CDF of the throughput for receiving devices in the star motif. 
\begin{corollary}
	\label{corollary3}
	Based on the distribution of the SIR, the CDF of the throughput for receivers in the star motif can be derived as follows:
	\begin{align}
	\label{f28star}
	\mathbb P(R_{j}<R)=1- \int_{0}^{\infty}\mathcal{L}^{\emph{\textrm{star}}}_{j\emph{\textrm{-inter}}}\left(\frac{2^{\frac{2R}{w_{1}}}-1}{P_{t}}r_{1}^{\alpha}\right)\mathcal{L}^{\emph{\textrm{star}}}_{j\emph{\textrm{-intra}}}\left(\frac{2^{\frac{2R}{w_{1}}}-1}{P_{t}}r_{1}^{\alpha}\right)f_{R_{1}}(r_{1})dr_{1}.
	\end{align}%
\end{corollary}

By using the relationship between the CDF and the expected value, we are able to obtain the expected throughput of receiving devices in the star motif in the following corollary. 

\begin{corollary}
	\label{lemma5star}
	The expected throughput for receivers in the star motif is given by:
	\begin{align}
	\label{f29star}
	E_{\emph{\textrm{star}}}&=\int_{0}^{\infty}(1-\mathbb P(R_{j}<R))dR \nonumber \\
	&= \int_{0}^{\infty} \int_{0}^{\infty}\mathcal{L}^{\emph{\textrm{star}}}_{j\emph{\textrm{-inter}}}\left(\frac{2^{\frac{2R}{w_{1}}}-1}{P_{t}}r_{1}^{\alpha}\right)\mathcal{L}^{\emph{\textrm{star}}}_{j\emph{\textrm{-intra}}}\left(\frac{2^{\frac{2R}{w_{1}}}-1}{P_{t}}r_{1}^{\alpha}\right)f_{R_{1}}(r_{1})dr_{1}dR.
	\end{align}%
\end{corollary}
\vspace{-0.6cm}
\subsection{Expected Throughput for Non-Seeding Devices in the Chain Motif}
\begin{figure}[!t]
	\centering
	\includegraphics[width=2.8in,height=2.1in]{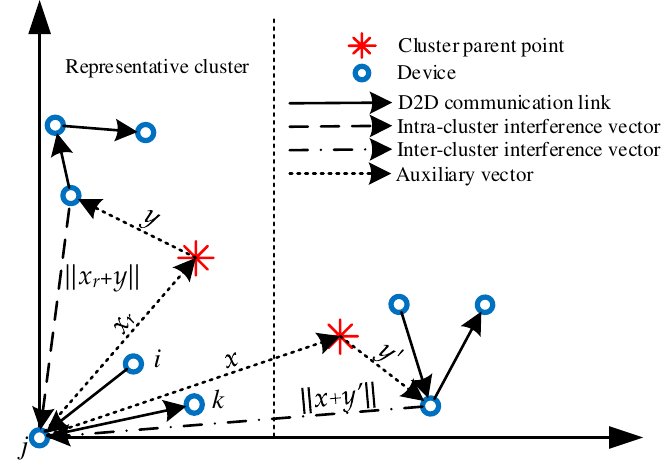}
	\DeclareGraphicsExtensions.
	\vspace{-0.1in}
	\caption{Illustration of the inter- and intra-cluster interference to the representative chain motif.}
	\label{Illustrate3}
	\vspace{-0.4in}
\end{figure}

\vspace{-0.2cm}
As illustrated in Fig. \ref{Illustrate3}, we also choose an arbitrary \textit{representative chain motif $M_{c}$} in cluster $C_{r}$. 
In motif $M_{c}$, the seeding device $i$ first transmits its data to $j$, which is at the origin and will subsequently propagate the data to device $k$. 
Similar to the receiving devices in the star motif, device $k$ in a chain motif also accesses its content from the transmitter at the origin.
Therefore, the Laplace transforms of the intra- and inter-cluster interference given by (\ref{intra-star}) and (\ref{inter-star}), and the probability in (\ref{f3201star}) can be directly applied to the D2D link between device $j$ and $k$. 
In contrast, due to the different assigned bandwidth, the CDF and the expected throughput must be re-calculated. 
Moreover, unlike device $k$, device $j$ located at the origin has different distance distribution to the serving device and the devices that generate interference from receiving nodes in the star motifs. 
Hence, to obtain the expression of the throughput, we need to derive the Laplace transforms of the intra- and inter-cluster interference for receiving device $j$ in the chain motif. 

\begin{lemma}
	\label{lemmathird}
	For the typical device $j$ in the chain motif, the Laplace transform of the intra-cluster interference, conditioned on the distance from the typical
	point to the parent point $s_{r}=||x_{r}||$, is given by 
	\begin{align}
	\label{f14}
	\mathcal{L}^{\emph{\textrm{chain}}}_{j\emph{\textrm{-intra}}}(s|s_{r})=&\left(\int_{0}^{\infty}\frac{1}{1+sP_{t}v_{3}^{-\alpha}}f_{V_{3}}(v_{3}|s_{r})dv_{3}\mathbb P_{S,S}+1-\mathbb P_{S,S}\right)^{N_{m}-1},
	\end{align}%
	where $f_{V_3}(v_{3}|s_{r})=f_{D}(v_{3},s_{r};\sigma^{2})$, $v_{3}\geq0$.
	\begin{proof}[Proof:\nopunct]
		See Appendix \ref{Appendix D}.
	\end{proof}
\end{lemma}

\begin{lemma}
	\label{lemmafourth}
	For the device $j$ in the chain motif, the Laplace transform of the inter-cluster interference is given by 
	\begin{align} 
	\label{f16}
	\mathcal{L}^{\emph{\textrm{chain}}}_{j\emph{\textrm{-inter}}}(s)\!\!=\!\!\exp\left(-\lambda_{p} 2\pi \int_{0}^{\infty}\left(1-\left(\int_{0}^{\infty}\frac{1}{1+sP_{t}v_{4}^{-\alpha}}f_{V_{4}}(v_{4}|t)dv_{4}\mathbb P_{S,S}+1-\mathbb P_{S,S}\right)^{N_{m}}\right)tdt\right),
	\end{align}%
	where $f_{V_{4}}(v_{4}|t)=f_{D}(v_{4},t;\sigma^{2})$, $v_{4}\geq0$.
	\begin{proof}[Proof:\nopunct]
		See Appendix \ref{Appendix E}.
	\end{proof}
\end{lemma}


As the D2D link between devices $i$ and $j$ and the link between $j$ and $k$ share the same bandwidth and the locations of $j$ and $k$ follow the same point process, the interference encountered by these two devices $j$ and $k$ can be correlated \cite{haenggi2012stochastic}. In the following theorem, we consider such interference correlation and calculate the outage probability for the chain motif. 
\begin{theorem}
	\label{lemma4}Taking into the interference correlation between two devices $j$ and $k$, 
	the outage probability of the chain motif can be given by:
	{\small\begin{align}
	\label{lemma41}
	\mathbb P^{\emph{\textrm{chain}}}_{\emph{\textrm{outage}}}(\delta_{\emph{\textrm{th}}})
	=&\int_{\R^{2}}\int_{\R^{2}}\int_{\R^{2}}1\!\!-\!\!\exp\Bigg(\!\!-\!\!\lambda_{p}\int_{\R^{2}}\Bigg[1\!\!-\!\!\Big(\int_{\R^{2}}\left[1\!\!-\!\!p\!\!+\!\!p
	\mathcal{L}_{h}\left(\frac{\delta_{\textrm{th}}}{||x_{r}\!\!+\!\!y_{i}||^{-\alpha}}(||x\!\!+\!\!y||)^{-\alpha}\right)\right] \nonumber \\  &\left[1\!\!-\!\!p\!\!+\!\!p\mathcal{L}_{h}\left(\frac{\delta_{\textrm{th}}}{||(x_{r}\!\!+\!\!y_{k})||^{-\alpha}}(||x\!\!+\!\!y\!\!-\!\!(x_{r}\!\!+\!\!y_{k})||)^{-\alpha}\right)\right]  \frac{1}{2\pi\sigma^{2}} \exp\left(\frac{-||y||^{2}}{2\sigma^{2}}\right)dy
	\Big)\!\Bigg]dx\!\Bigg)dy_{i}dy_{k}dx_{r},
	\end{align}}%
	where $p=\frac{N_{m} \times  \mathbb P_{S,S}}{N}$, and $\mathcal{L}_{h}(\cdot)$ represents the Laplace transform of the fading distribution.
	\begin{proof}[Proof:\nopunct]
		See Appendix \ref{AppendixOutageChainProof}.
	\end{proof} 
\end{theorem}
Even though Theorem \ref{lemma4} characterizes the interference in presence of correlation, one can see from (\ref{lemma41}) that this expression can be computationally complex to derive. 
As such, in order to obtain a more tractable result, hereinafter, we consider a special case where there is no correlation between the interference encountered by devices $j$ and $k$. 
Such a case has been considered in recent works such as \cite{dai2017analytical} and \cite{al2016stochastic}. Next, the simplified outage probability for the chain motifs is calculated. 

\begin{corollary}
	\label{Corollary7}
	When there is no correlation between the interference experienced by two receivers in the chain motif, the outage probability of chain motifs is 
	\begin{align}
	\label{Corollary7results}
	\mathbb P^{\emph{\textrm{chain}}}_{\emph{\textrm{outage}}}(\delta_{\emph{\textrm{th}}})
	&=1- \left( \int_{0}^{\infty}\mathcal{L}^{\emph{\textrm{star}}}_{j\emph{\textrm{-inter}}}\left(\frac{\delta_{\emph{\textrm{th}}}}{P_{t}}r_{1}^{\alpha}\right)\mathcal{L}^{\emph{\textrm{star}}}_{j\emph{\textrm{-intra}}}\left(\frac{\delta_{\emph{\textrm{th}}}}{P_{t}}r_{1}^{\alpha}\right)f_{R_{1}}(r_{1})dr_{1}\right)\times \nonumber \\ &\left(\int_{0}^{\infty} \int_{0}^{\infty}\mathcal{L}^{\emph{\textrm{chain}}}_{j\emph{\textrm{-inter}}}\left(\frac{\delta_{\emph{\textrm{th}}}}{P_{t}}r_{2}^{\alpha}\right)\mathcal{L}^{\emph{\textrm{chain}}}_{j\emph{\textrm{-intra}}}\left(\frac{\delta_{\emph{\textrm{th}}}}{P_{t}}r_{2}^{\alpha}|s_{r}\right)f_{R_{2}}(r_{2}|s_{r})f_{S_{r}}(s_{r})dr_{2}ds_{r}\right),
	\end{align}%
	where $f_{R_{2}}(r_{2}|s_{r})=f_{D}(r_{2},s_{r};\sigma^{2})$, $r_{2}\geq 0$, and $f_{S_{r}}(s_{r})=f_{S}(s_{r};\sigma^{2})$, $s_{r}\geq 0$.
	\begin{proof}[Proof:\nopunct]
		See Appendix 
		For the D2D link from device $i$ to $j$, the probability of the D2D link from device $i$ to $j$ meeting the minimum SIR threshold is
		\begin{align}
		\label{f3201}
		\mathbb P^{\textrm{chain}}_{j}(\delta_{\textrm{th}})&=\E(\mathbb{P}\{\delta_{j}>\delta_{\textrm{th}}\})\nonumber \\
		& =\E_{R}\left[\mathbb{P}\left(h_{ij}>\frac{\delta_{\textrm{th}}}{P_{t}}(I^{x_{r}}_{j\textrm{-intra}}+I^{x_{r}}_{j\textrm{-inter}})||x_{r}+y_{j}||^{\alpha}\right)\right] \nonumber \\
		&=\E_{R}\left[\E\left[\exp\left(\frac{\delta_{\textrm{th}}}{P_{t}}(I^{x_{r}}_{j\textrm{-intra}}+I^{x_{r}}_{j\textrm{-inter}})||x_{r}+y_{j}||^{\alpha}\right)\right]\right]  \nonumber\\
		&\stackrel{(a)}{=}\int_{0}^{\infty} \int_{0}^{\infty}\mathcal{L}^{\textrm{chain}}_{j\textrm{-inter}}\left(\frac{\delta_{\textrm{th}}}{P_{t}}r_{2}^{\alpha}\right)\mathcal{L}^{\textrm{chain}}_{j\textrm{-intra}}\left(\frac{\delta_{\textrm{th}}}{P_{t}}r_{2}^{\alpha}|s_{r}\right)f_{R_{2}}(r_{2}|s_{r})f_{S_{r}}(s_{r})dr_{2}ds_{r}.
		\end{align}%
		In $(a)$, we first make a change of variables by setting $r_{2}=||x_{r}+y_{j}||$, and, then, we perform one de-conditioning in terms of $r_{2}$ and another de-conditioning for $s_{r}$, and finally convert the coordinates from Cartesian to polar. Based on the distance distribution,  we have $f_{R_{2}}(r_{2}|s_{r})=f_{D}(r_{2},s_{r};\sigma^{2})$ and $f_{S_{r}}(s_{r})=f_{S}(s_{r};\sigma^{2})$. Then, since the outage will not occur if and only if the SIR of both links exceed the threshold level, we can obtain (\ref{Corollary7results}).
	\end{proof} 
\end{corollary}
Similar to Corollary \ref{corollary3}, we can replace  $\delta_{\textrm{th}}$ with $2^{\frac{R}{w_{1}}}-1$ in (\ref{f3201star}) and (\ref{f3201}) to obtain the throughput CDF for the two D2D links in the chain motif.  
To calculate the throughput, we can observe that the chain motif is equivalent to a two-hop network. According to the work in \cite{sikora2006bandwidth}, if a non-seeding node is connected to the seeding node through a multi-hop link with length $l$, the achievable throughput of the non-seeding device is limited by $1/l$ times the minimum of all D2D link rates over the multi-hop communication network. Therefore, we can derive the expected throughput of non-seeding devices $j$ and $k$ in the following corollaries. 

\begin{corollary}
	\label{lemma5}
	Since device $j$ is directly connected to seeding node $i$, the expected throughput is 
	\begin{align}
	\label{f29}
	E_{\emph{\textrm{chain}}}&=\frac{1}{1}\min \Big(\int_{0}^{\infty}(1-(1-\mathbb P^{\emph{\textrm{chain}}}_{\textrm{j}}(2^{\frac{R}{w_{1}}}-1)))dR\Big) \nonumber \\
	&=\!\!\int_{0}^{\infty}\!\!\int_{0}^{\infty}\!\!\! \int_{0}^{\infty}\!\!\mathcal{L}^{\emph{\textrm{chain}}}_{j\emph{\textrm{-inter}}}\left(\!\frac{2^{\frac{R}{w_{1}}}\!-\!1}{P_{t}}r_{2}^{\alpha}\!\right)\mathcal{L}^{\emph{\textrm{chain}}}_{j\emph{\textrm{-intra}}}\left(\!\frac{2^{\frac{R}{w_{1}}}\!-\!1}{P_{t}}r_{2}^{\alpha}|s_{r}\!\right)\!f_{R_{2}}(r_{2}|s_{r})f_{S_{r}}(s_{r})dr_{2}ds_{r}dR.
	\end{align}
\end{corollary}
\begin{corollary}
	\label{lemma6}
	Since device $k$ connects to the seeding device $i$ via a two-hop link, the expected throughput is given by 
	\begin{align}
	\label{f291}
	&E_{\emph{\textrm{chain}}}'=\frac{1}{2}\min \Bigg(\int_{0}^{\infty}   \int_{0}^{\infty}\mathcal{L}^{\emph{\textrm{star}}}_{j\emph{\textrm{-inter}}}\left(\frac{2^{\frac{R}{w_{1}}}-1}{P_{t}}r_{1}^{\alpha}\right)\mathcal{L}^{\emph{\textrm{star}}}_{j\emph{\textrm{-intra}}}\left(\frac{2^{\frac{R}{w_{1}}}-1}{P_{t}}r_{1}^{\alpha}|s_{r}\right)f_{R_{1}}(r_{1})dr_{1} dR,E_{\emph{\textrm{chain}}} \Bigg).
	\end{align}%
\end{corollary}
\vspace{-0.6cm}
\subsection{Expected Throughput for Seeding Devices}
\vspace{-0.2cm}

From the work in \cite{afshang2016modeling}, conditioned on the distance between the parent point $x_{r} \in \Phi_{p}$ and a typical device at the origin, the distance between the device in cluster $C_{r}$ and the typical device follows a Rician distribution. 
Equivalently, if we set the condition as the distance $z=||x_{r}-b||$ between the parent point $x_{r}$ and the BS $b$, the distance $r_{3}$ between the devices in the cluster and the BS will also follow a Rician distribution $f_{R_{3}}(r_{3}|z)=f_{D}(r_{3},z;\sigma^{2})$, $z, r_{3}>0$. 
Similar to (\ref{f28star}), we change the coordinates and replace the variables, and, then, the CDF of throughput will be: 
\begin{align}
\label{cellular devices}
P(R_{k}<R)&=1-\int_{\R^{2}} \int_{\R^{2}} -\exp \left(\frac{2^{\frac{R}{w_{2}}}-1}{P_{b}} P_{n}||x_{r}+y-b||^{\alpha}\right)f_{Y}(y)f_{X}(x_{r})dx_{r}dy \nonumber\\
&=1 -  \int_{0}^{\infty} \int_{0}^{\infty}-\exp \left(\frac{2^{\frac{R}{w_{2}}}-1}{P_{b}} P_{n}r_{3}^{\alpha}\right)f_{R_{3}}(r_{3}|z)f_{Z}(z)dr_{3}dz, 
\end{align}
where $r_{3} = ||x_{r}+y-b||$. 

According to the distribution of parent points, $z$ can be interpreted as the distance between the BS and a point which is uniformly distributed in the area. 
Without loss of generality, we assume that the BS is at the center of a square area of side $2L$ that contains all devices. Hence, the CDF expression for the distance $z$ will be:
\begin{align}
F_{Z}(z) = \frac{1}{4L^{2}}\left\{ \begin{array}{cc} 
\pi z^{2},&0\leq z \leq L, \\
\pi z^{2}-4\left(z^{2}\arccos\left(\frac{L}{z}\right)-L\sqrt{z^{2}-L^{2}}\right),  & L\leq z \leq \sqrt{2}L,  \\
1, &\sqrt{2}L\leq z.  \\
\end{array} \right.
\end{align}

We can then find the corresponding PDF as 
\begin{align}
f_{Z}(z) = \frac{1}{2L^{2}}\left\{ \begin{array}{cc} 
\pi z,&0\leq z \leq L, \\
\pi z- 4 z \arccos\left(\frac{L}{z}\right),  & L\leq z \leq \sqrt{2}L,  \\
0, & \text{otherwise}.  \\
\end{array} \right.
\end{align}

Based on the relationship between the SIR and the expected throughput, we can obtain the expression of the throughput in the next Lemma. 
\begin{lemma} For seeding devices within the D2D network, the expected throughput is given by 
	\begin{align}
	\label{cellulardevices}
	E_{\emph{\textrm{seeding}}}&=\int_{0}^{\infty}\int_{0}^{\infty} \int_{0}^{\infty}\frac{2^{\frac{R}{w_{2}}}-1}{P_{b}} P_{n}r_{3}^{\alpha}f_{R_{3}}(r_{3}|z)f_{Z}(z)dr_{3}dz dR.
	\end{align}%
\end{lemma}

\subsection{Average Throughput per Device in the D2D Network}
After obtaining the throughput for each device and the possibility of a three-node motif being either a star or a chain motif, the expected average throughput per device in the D2D network can be derived next.

\begin{corollary}
	\label{corollary11}
	Based on (\ref{f29star}), (\ref{f29}), (\ref{f291}), as well as (\ref{cellulardevices}), and the possibilities of a three-node motif being either a star or a chain motif, the average throughput per device will be: 
	\begin{align}
	\label{average1}
	E_{R}&=\frac{2c^{\emph{\textrm{star}}}_{\emph{\textrm{o}}}E_{\emph{\textrm{star}}} + c^{\emph{\textrm{star}}}_{\emph{\textrm{o}}}E_{\emph{\textrm{seeding}}} + c^{\emph{\textrm{chain}}}_{\emph{\textrm{o}}}E_{\emph{\textrm{chain}}} + c^{\emph{\textrm{chain}}}_{\emph{\textrm{o}}}E_{\emph{\textrm{chain}}}' + c^{\emph{\textrm{chain}}}_{\emph{\textrm{o}}}E_{\emph{\textrm{seeding}}} + E_{\emph{\textrm{seeding}}}}{N} \nonumber \\
	&=\frac{2c^{\emph{\textrm{star}}}_{\emph{\textrm{o}}}E_{\emph{\textrm{star}}} + c^{\emph{\textrm{chain}}}_{\emph{\textrm{o}}}E_{\emph{\textrm{chain}}} + c^{\emph{\textrm{chain}}}_{\emph{\textrm{o}}}E_{\emph{\textrm{chain}}}' + (c_{\emph{\textrm{o}}}+1)
		E_{\emph{\textrm{seeding}}}}{N}.
	\end{align}
\end{corollary}
We can observe that 
the average throughput per device in the content dissemination network is dependent on the same factors on which the $Z$-scores of both the star and chain motifs depend, namely, the
distribution variance, the number of devices, and the maximum allowable distance. 
Thus, both the system throughput and $Z$-score for motifs are functions of the same network parameters, and, hence, the BS can calculate the $Z$-scores of chain and star motifs and the system throughput corresponding to the same settings.
These numerical results on the $Z$-score and system throughput can help understand how the dissemination performance changes when the occurrence of different motifs varies.
Furthermore, the numerical results can also help identify $Z$-score regions, which map to the optimal content dissemination performance, for both chain and star motifs, as will also be evident from the subsequent simulation results.	\vspace{-0.2in}

\begin{table}[!t]
	\large
	\begin{center}
		\caption{\small Simulation parameters.}
		\vspace{-0.2cm}
		\label{parameters}
		\footnotesize
		\resizebox{9cm}{!}{
			\begin{tabular}{|c|c|c|c}
				\hline
				\textbf{Parameter} & Definition  & \textbf{Value} \\ \hline
				$P_{b}$ & Transmission power of BS  & $46$~dBm \cite{xu2014resource}\\ \hline
				$P_{t}$ & Transmission power of devices & $23$~dBm \cite{xu2014resource}\\ \hline
				$N$ & Number of devices in each cluster  &  $25,50,100$ \cite{afshang2016modeling}\\ \hline
				$\sigma^{2}$ & Distribution variance & $50,100,150$ \cite{afshang2016modeling} \\ \hline
				$w$ &Total system bandwidth & $20$~MHz \cite{xu2014resource} \\ \hline
				$\beta$ & Bandwidth parameter & $0.6$  \\ \hline
				$\alpha$ & Path loss exponent &  $4$ \cite{afshang2016fundamentals}\\ \hline
				$\delta_{\text{th}}$ & Minimum SIR requirement & $0$~dB \cite{afshang2016fundamentals}\\ \hline
				\slash &Noise spectral density & $-174$~dBm/Hz \\ \hline 
				\end{tabular}}
	\end{center}
	\vspace{-0.5in}
\end{table}

\section{Simulation Results}
	\vspace{-0.1in}

Next, we present rigorous simulation results to validate the analytical derivations and show the relationship between the occurrence of motifs and the content dissemination performance in D2D-enabled cellular networks.
In particular, we consider that wireless devices are distributed within a square area of $1$~km $\times$ $1$~km. 
Also, for each D2D link, the transmitter is chosen randomly. 
To observe the impacts of different network topology parameters on the occurrence of motifs and the system throughput, we change the distribution variance and the cluster density. Simulation parameters are summarized in Table \ref{parameters}.


\begin{figure}
	\centering	
	\begin{minipage}{0.45\textwidth}
		\centering
		\includegraphics[width=1\linewidth]{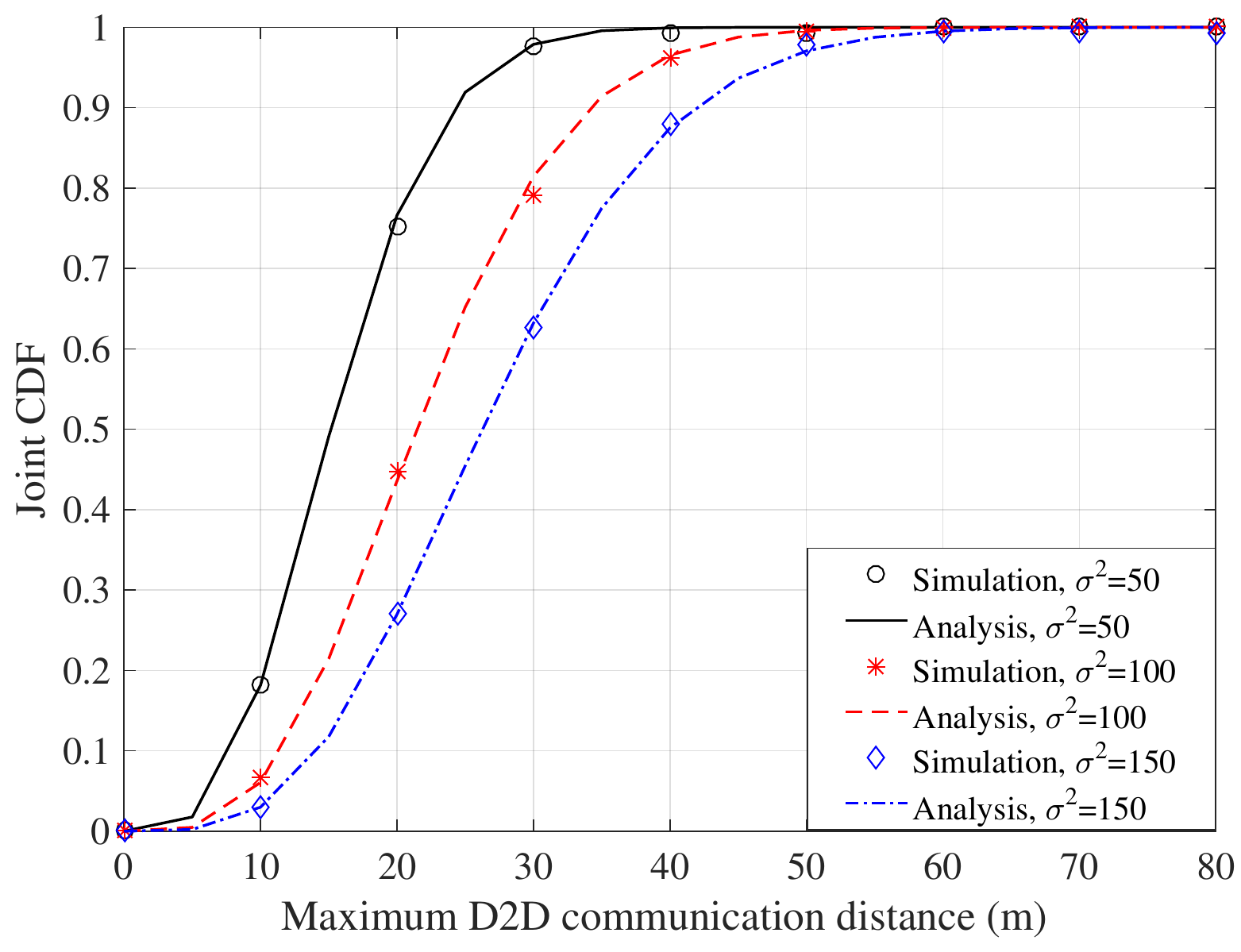}
		\vspace{-0.2in}
		\caption{Joint CDF versus threshold  $s_{\text{th}}$.}
		\label{simulation1}
	\end{minipage}
	\begin{minipage}{0.43\textwidth}
		\centering
		\includegraphics[width=1\linewidth]{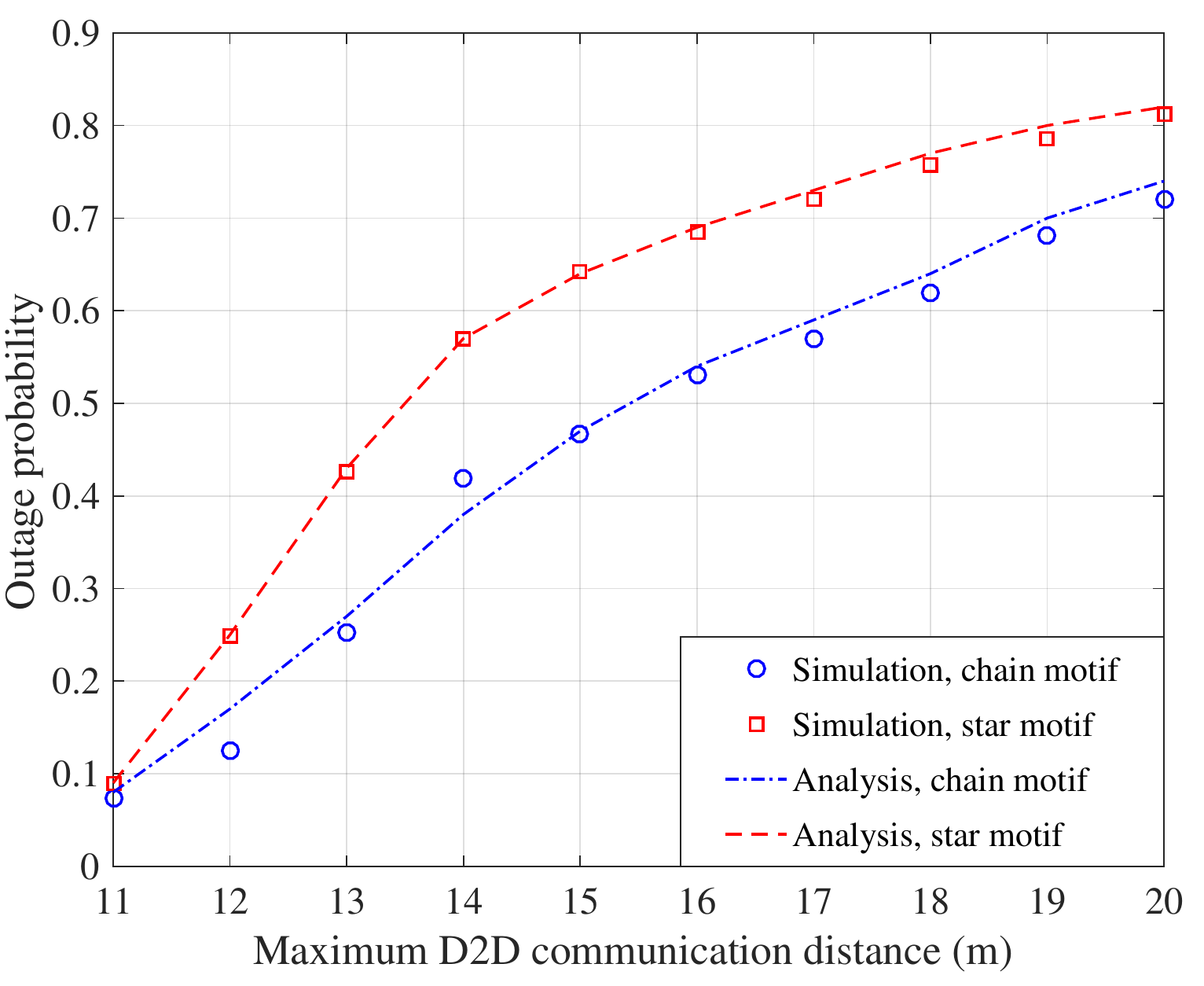}
		\vspace{-0.2in}
		\caption{Outage probability versus threshold $s_{\text{th}}$.}
		\label{simulation11}
	\end{minipage}
	\vspace{-0.4in}
\end{figure}


Fig. \ref{simulation1} shows the joint CDF $\mathbb P_{S,S}$ of a group of three devices being a three-node motif versus the communication distance $s_{\text{th}}$ for systems with $\lambda_{p}=20$ $\text{cluster}/\text{km}^{2}$ and $\sigma^{2}=50,100,150$.
As shown in Fig. \ref{simulation1}, we can observe that the simulation results corroborate the analytical results in Theorem \ref{lemma1}. 
Moreover, as $s_{\text{th}}$ increases, $\mathbb P_{S,S}$ increases. 
But, the increase varies differently for different D2D systems. 
In particular, to achieve $\mathbb P_{S,S}=1$, the $s_{\text{th}}$ for the D2D system with $\sigma^{2}=50$ should be around $40$~m, while the ones of systems with $\sigma^{2}=100,150$ are, respectively, close to $50$~m and $60$~m. 
This is due to the fact that, in the system with a smaller distribution variance, devices are more likely to form three-node motifs. 
Furthermore, for a fixed distance threshold $s_{\text{th}}$, we can observe that the system with a lower variance will lead to a higher probability $\mathbb P_{S,S}$. For example, when $s_{\text{th}}=20$~m, the values of $\mathbb P_{S,S}$ are approximately $0.75$, $0.45$, and $0.28$ for the systems with $\sigma^{2}=50, 100, 150$.


Fig. \ref{simulation11} shows the outage probability, for chain and star motifs, versus the D2D communication distance $s_{\text{th}}$
with $\delta_{\text{th}}=0$~dB, $\sigma^{2}=100$, and $N = 25$ in each cluster by using Theorem \ref{lemmastar} and Corollary \ref{Corollary7}. 
As illustrated in Fig. \ref{simulation11}, the simulation results of the star motifs validate the analytical results in Theorem \ref{lemmastar}.
For the chain motifs, there are some mismatch between the simulation and analytical results, which can be explained that we did not consider interference correlation in Corollary \ref{Corollary7}.
Moreover, we can observe that, as $s_{\text{th}}$ increases, the outage probabilities for both motifs will increase. 
This is due to the fact that as the maximum distance increases, the D2D system is more likely to have more motifs, leading to more interference and thereby a lower SIR.

\begin{figure}
	\centering	
	\begin{minipage}{0.42\textwidth}
		\centering
		\includegraphics[width=1\linewidth]{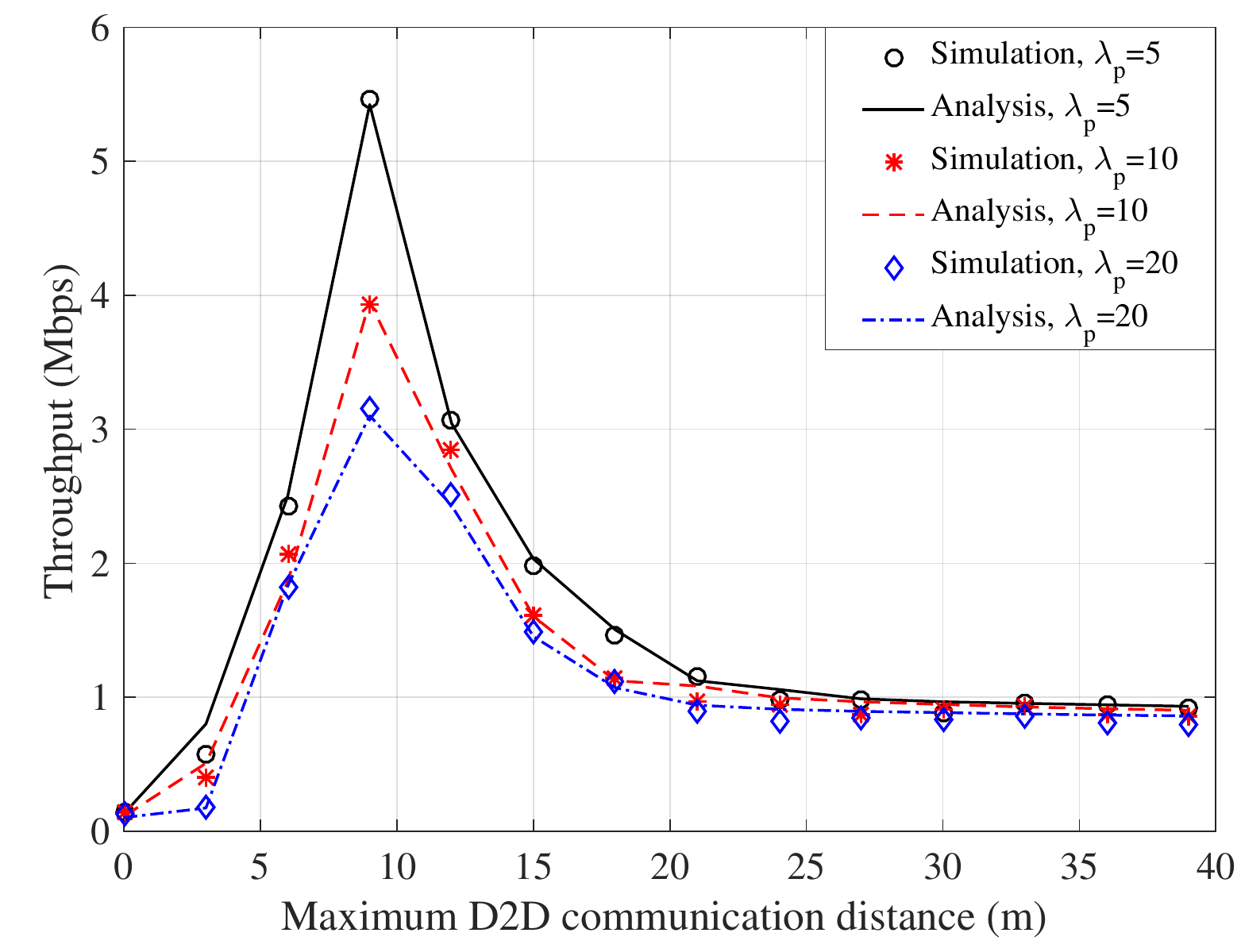}
		\caption{Average data per device versus the distance threshold $s_{\text{th}}$.}
		\label{simulation2}
	\end{minipage}
	\begin{minipage}{0.48\textwidth}
		\centering
		\includegraphics[width=1\linewidth]{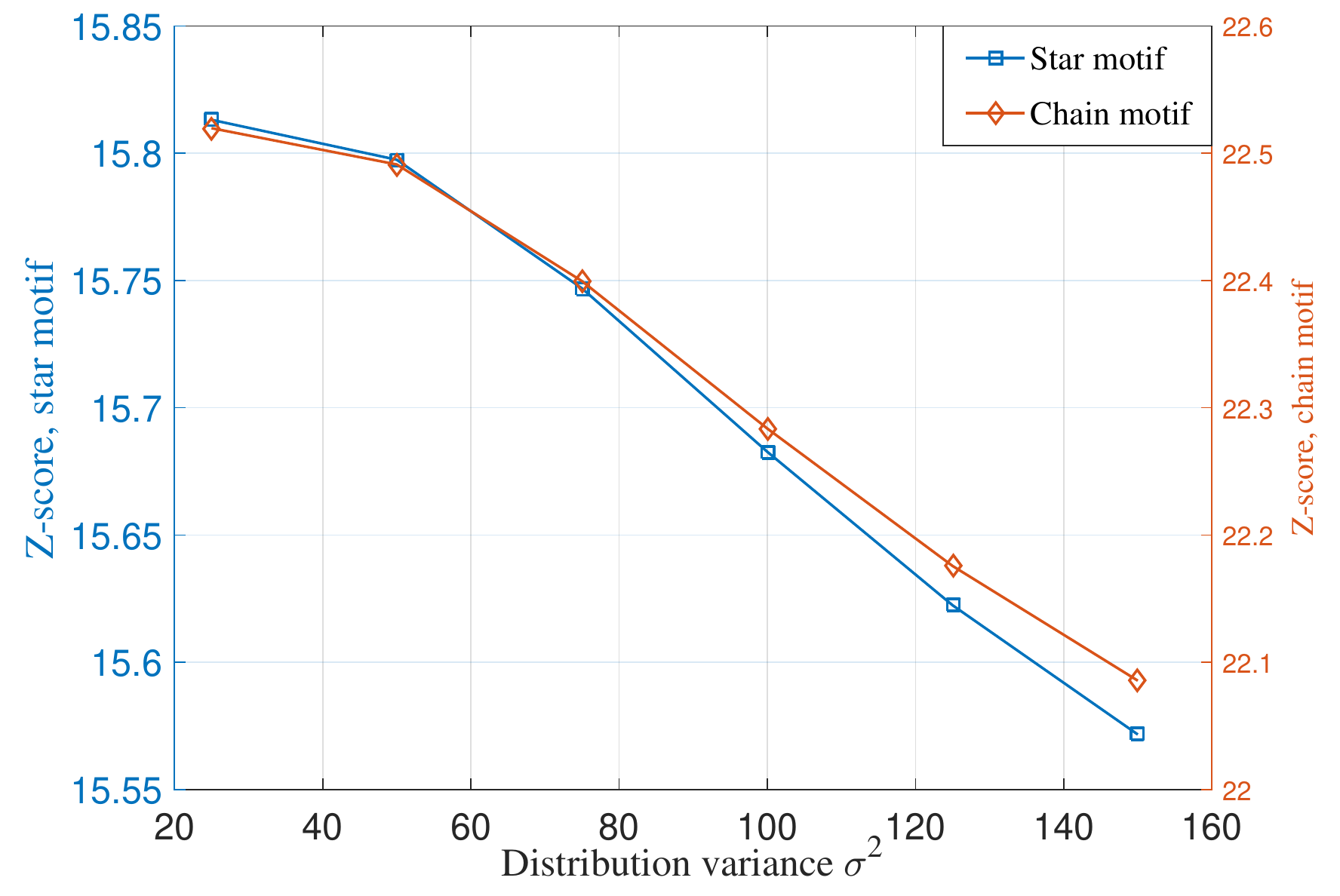}
		\caption{$Z$-scores for the chain and star motifs versus the distribution variance $\sigma^{2}$.}
		\label{simulation61}
	\end{minipage}
	\vspace{-0.2in}
\end{figure}

Fig. \ref{simulation2} shows the average throughput per device when $\sigma^{2}=100$, $\lambda_{p}=5,10,20$ $\text{cluster}/\text{km}^{2}$, and $N=50$. 
In Fig. \ref{simulation2}, we can observe that the simulation results of average throughout corroborate the analytical derivations in Corollary \ref{corollary11}. 
Moreover, Fig. \ref{simulation2} shows that the cluster density $\lambda_{p}$ plays an important role in determining the content dissemination performance. 
In particular, when $s_{\text{th}}$ is small, the system with a smaller cluster density $\lambda_{p}$ will have a higher throughput. 
However, when $s_{\text{th}}$ is above $30$~m, the throughput of D2D systems with lower cluster densities is almost equal to the counterpart for systems with higher cluster densities.
In fact, when $s_{\text{th}}$ is very small, both motifs barely exist in the networks and, thus, D2D links will experience less intra-cluster interference. 
In this case, the system performance mainly depends on the inter-cluster interference. 
Therefore, a system with a lower cluster density tends to have a better performance than the one with a higher density.  
However, when $s_{\text{th}}$ is larger, more and more three-node motifs will appear in the network, increasing both inter- and intra-cluster interference. 
From the analytical results in Section \uppercase\expandafter{\romannumeral4}, we can observe that, when the $s_{\text{th}}$, $N$, and $\sigma^{2}$ are fixed, intra-cluster interference will be the same for systems with different $\lambda_{p}$. 
But, due to the different cluster densities, the inter-cluster interference will be different. 
We can observe that, when $s_{\text{th}}$ increases to $25$~m, the average throughput is almost equal for all $\lambda_p$. This is can be explained that, in such a case, the intra-cluster interference has a dominant effect  on the system performance.

Fig. \ref{simulation61} presents the $Z$-scores for the chain and star motifs versus the distribution variance $\sigma^{2}$. 
We can observe that, as the distribution variance $\sigma^{2}$ increases, the $Z$-scores for both chain and star motifs will decrease. 
The reason is that, for a fixed $s_{\text{th}}$, as the variance $\sigma^{2}$ increases, both motifs are less likely to occur in the network. 
Moreover, we can observe that $Z$-scores of chain motifs are higher than the counterparts for star motifs, indicating that chain motifs are more likely to occur in network. 
It can be explained that star motifs need more restrictive requirements, i.e., two different D2D links sharing the same seeding device.
By combining the results of Fig. \ref{simulation2} and Fig. \ref{simulation61} with the analytical derivations in Section \ref{$Z$-score} and Section \ref{throughput}, we can see that both the average throughput per device and the $Z$-scores for the motifs are dependent on the same factors, i.e., $\sigma^{2}$ and $s_{\text{th}}$. 

\begin{figure}[t!] 
	\centering
	\subfloat{%
		\includegraphics[width=0.46\textwidth]{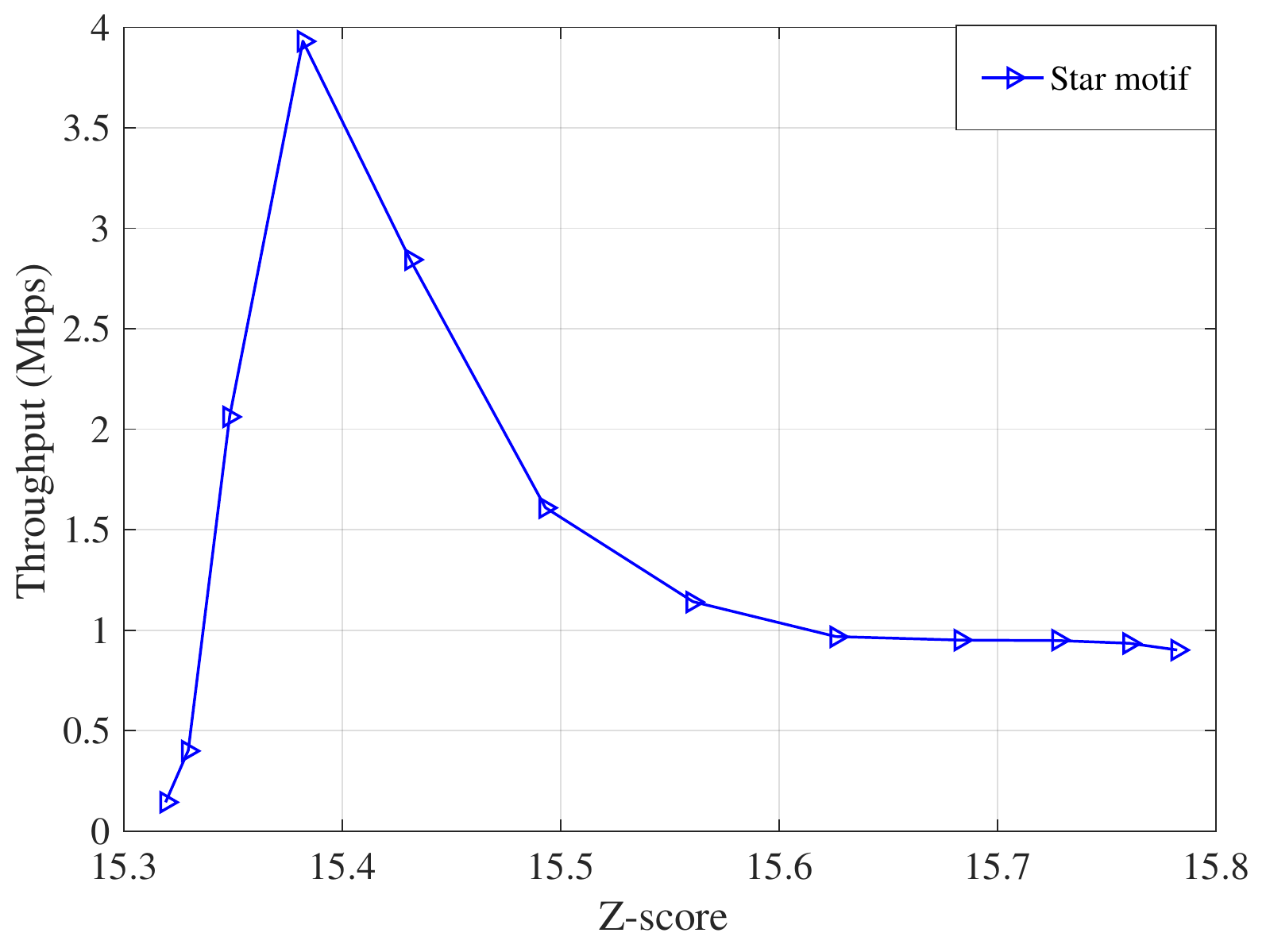}%
		\label{fig:a}%
	}	
	\subfloat{%
		\includegraphics[width=0.46\textwidth]{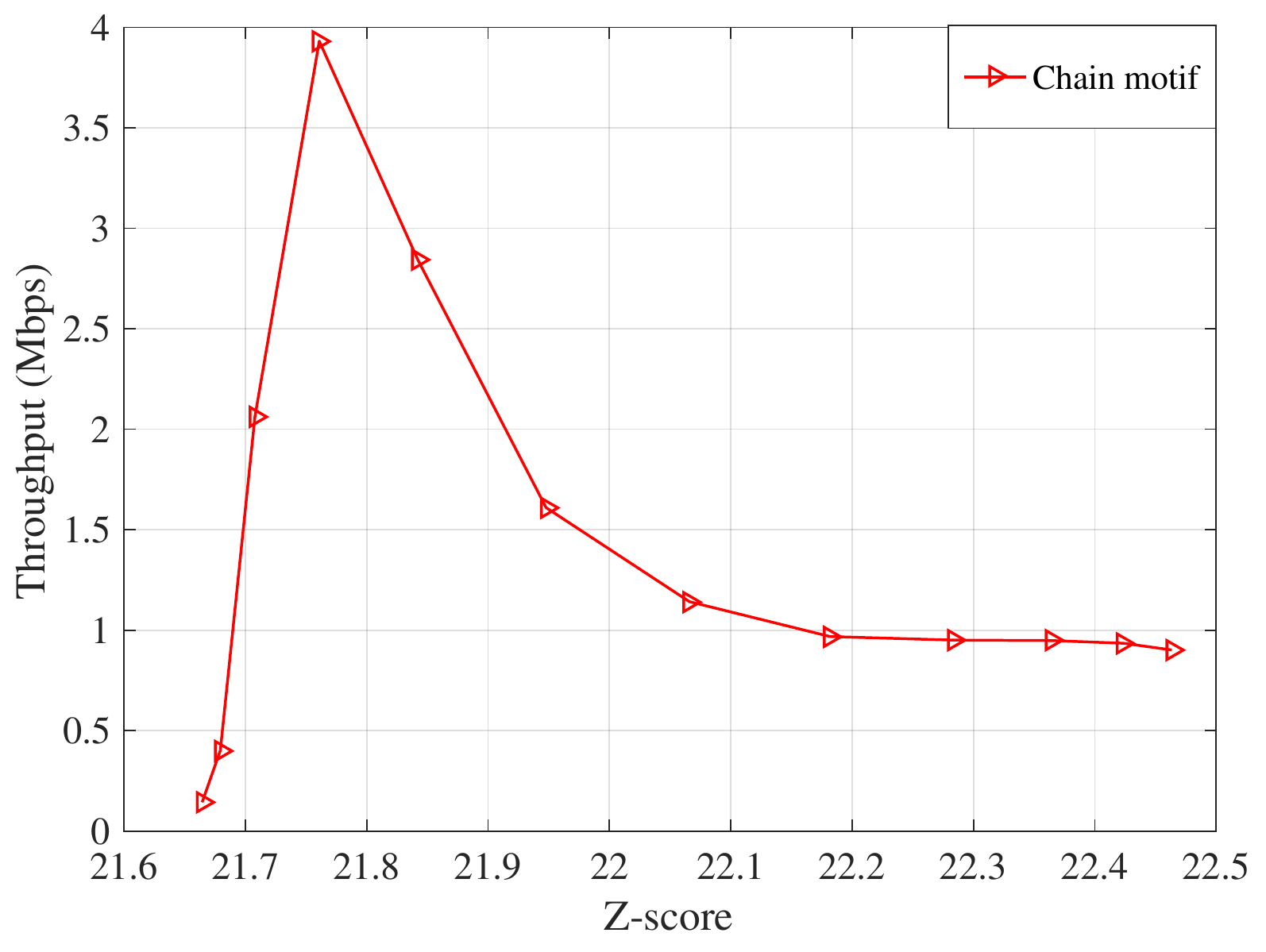}%
		\label{fig:b}%
	}	
	\vspace{-0.2in}
	\caption{Average throughput per device versus $Z$-scores for the star and chain motifs.}
	\label{simulation6}
	\vspace{-0.2in}
\end{figure}

Fig. \ref{simulation6} shows the average throughput per device versus the $Z$-scores for the chain and star motifs. 
Here, we consider a network with $\lambda_p=10$ $\text{clusters}/\text{km}^{2}$, $\sigma^2=100$ and $N=50$.
Moreover, we change the value of $s_{\text{th}}$ so that we can change the $Z$-scores for both motifs as well as the system throughput. 
From Fig. \ref{simulation6}, we can see that, for both motifs, the average throughput is a concave function with respect to the $Z$-scores for both motifs. 
Moreover, there exists $Z$-score regions for both chain and star motifs that lead to a higher throughput than points in other $Z$-score regions. 
This stems from the tradeoff between received signal power over the D2D communication links and the interference introduced from the increased number of D2D links. 
In addition, the optimal $Z$-score regions will be different for star and chain motifs. 
For example, for chain motifs, the $Z$-score region where the network can achieve at least $2$~Mbps throughput is from $21.71$ to $21.92$, and the counterpart for the star motif is from $15.35$ to $15.47$.  

\begin{figure}
	\centering
	\includegraphics[width=.46\textwidth]{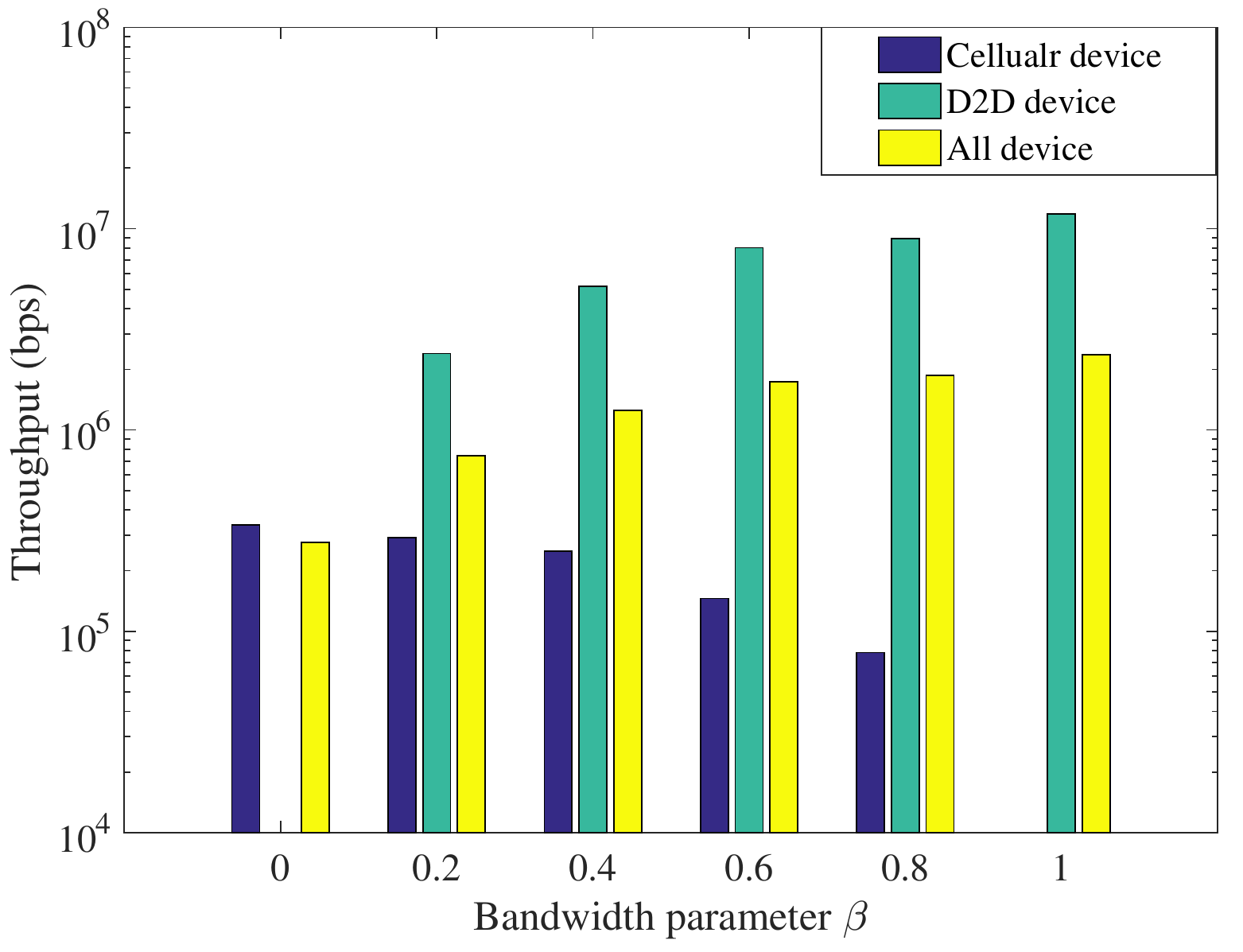}
	\vspace{-0.1in}
	\caption{Average throughput of D2D, cellular, and all devices versus bandwidth parameter $\beta$.}
	\label{datacellulard2d}
	\vspace{-0.2in}
\end{figure}

Fig. \ref{datacellulard2d} shows the average throughput for D2D and cellular devices under different bandwidth parameter $\beta$, when $\sigma^{2}=100$, $s_{\text{th}}=15$~m, and $\lambda_{p}=10$ $\text{clusters}/\text{km}^{2}$.
We can observe that as the bandwidth parameter increases, the average D2D throughput will increase while the throughput for cellular devices decreases. 
We can conclude that the system performance is mainly determined by the average throughput of D2D devices. 
Therefore, when designing a D2D system, the bandwidth parameter $\beta$ should be properly chosen so that the D2D-enabled cellular system can achieve the target performance. 

From the simulation results shown in Figs. \ref{simulation2}-\ref{datacellulard2d}, we can observe that, as long as the operator knows the $Z$-scores of the chain and star motifs appearing in the D2D communication networks, the achievable content dissemination performance, captured by the average throughput, can be predictable.
The predictable throughput can enable the operator to find the optimal $Z$-score regions which can achieve the best dissemination performance. 
Hence, when using D2D communications for content dissemination, the operators can identify the proper clusters to achieve a targeted performance by verifying whether the $Z$-score is in the optimal region.
Furthermore, since the $Z$-score is directly linked to the number of occurrences of chain and star motifs in the network, we can also determine the number of seeding devices in each identified cluster.
As such, based on the relationship between motif occurrence and content dissemination performance, the operators can decide on how to identify proper clusters to leverage D2D communications and further determine the number of seeding devices in each identified cluster for the optimal content dissemination policy.
Note that we can also use our proposed framework to study the content dissemination policy for general motifs with more nodes. 
That is, we can derive the corresponding $Z$-score for the motif of interest by calculating the expected number of occurrence in the D2D network and random graph and the standard derivation of occurrence in the random graph. 
Similar to the analysis for the three-node motifs, we can perform interference analysis and calculate the outage probability and the system throughput. 
Therefore, we can identify the relationship between the occurrence of the general motif and the system performance by numerical results and design a suitable content dissemination policy for the network operator.

\section{Conclusion}
In this paper, we have proposed a novel D2D network analysis framework
to explore frequent communication patterns across D2D users,
known as spatial motifs, to optimize content dissemination in emerging cellular networks. 
In particular, we have utilized the stochastic geometry to derive the analytical results of the frequency of occurrence for two common types of motifs composed of three devices, and the performance metrics for each motif in the D2D-enabled clustered network. 
Simulation results have corroborated the analytical derivations and shown the impact of system topologies on the occurrence of motif and the average throughput. 
More importantly, the simulation results have shed light on the relationship between the average throughput and the motifs appearing in the D2D network.
The derived relationship can be used by network operators to obtain the statistical significance regions for different motifs observed in the network, and to determine the content dissemination strategy -- e.g., allocating bandwidth to D2D links, determining number of seeding nodes -- in future cellular networks.
Therefore, by implementing the proposed framework, the operator can obtain
guidelines for designing effective content dissemination strategies in D2D-enabled cellular networks.

\appendix
\vspace{-0.1cm}
\subsection{Proof of Theorem \ref{lemma1}}
\label{appendixtotheorem1}
To calculate the probability of both $s_{1}=||m||$ and $s_{2}=||n||$ being smaller than the distance threshold $s_{\textrm{th}}$, we first calculate the correlation coefficient existing between $m$ and $n$ as 
{\small
	\begin{align}
	\label{f5}
	\rho=\frac{\textrm{cov}(m,n)}{\epsilon_{m}\epsilon_{n}}=\frac{\textrm{cov}(m,n)}{\sqrt{\textrm{cov}(m,m)}\sqrt{\textrm{cov}(n,n)}}=\frac{1}{2}.
	\end{align}}%
where $\epsilon_{m}$ and $\epsilon_{n}$ are the standard deviations for $m$ and $n$, respectively. 
Then, due to the fact that both $||m||$ and $||n||$ follow Rayleigh distribution, we are able to have the joint cumulative density function (CDF) as \cite{simon2007probability}
\begin{align}
\label{f6}
\mathbb P_{S,S}(s_{1},s_{2})&=\frac{3}{4}\sum_{k=0}^{\infty}\left(\frac{1}{4}\right)^{k}\frac{\gamma\left(1+k,\frac{ s_{1}^{2}}{3\sigma^{2}}\right)\gamma\left(1+k,\frac{ s_{2}^{2}}{3
		\sigma^{2}}\right)}{k!\varGamma(k+1)}, s_{1}\geq 0, s_{2}\geq 0,
\end{align}
where 
$\varGamma(\cdot)$ represents to the complete gamma function, defined as $\varGamma(a)=(a-1)!$. After replacing $s_{1}$ and $s_{2}$ with $s_{\textrm{th}}$ and simplifications, we can obtain the joint CDF shown in (\ref{f7}).
\subsection{Proof of Lemma \ref{lemmafirst}}
\vspace{-0.1cm}
We conduct the Laplace transform of the intra-cluster interference of device $j$ as 
\label{Appendix B}
{\small
\begin{align}
\label{f251}
\mathcal{L}^{\textrm{star}}_{j\text{-intra}}(s)&\stackrel{(a)}{=}\E\left[\exp\left(-s\sum_{y\in \mathcal{T}_{x_{r}} \setminus y_{i}}P_{t}h_{yj}||y_{j}-y||^{-\alpha}\right)\right] \nonumber \\
&=\E_{\mathcal{T}_{x_{r}}}\left[\prod_{y\in \mathcal{T}_{x_{r}} \setminus y_{i}}\E_{h_{yj}}[\exp(-sP_{t}h_{yi}||y-y_{j}||^{-\alpha})]\right] \nonumber \\
&\stackrel{(b)}{=}\E_{ \mathcal{T}_{x_{r}}}\left[\prod_{y\in \mathcal{T}_{x_{r}} \setminus y_{i}}\frac{1}{1+sP_{t}||y-y_{j}||^{-\alpha}}\right] \nonumber \\
&\stackrel{(c)}{=}\sum_{k=0}^{N_{m}-1}\left(\E\left[\frac{1}{1+sP_{t}||y-y_{j}||^{-\alpha}}\right]\right)^{k}\times \underbrace{{{N_{m}-1}\choose{k}}(\mathbb P_{S,S})^{k}(1-\mathbb P_{S,S})^{N_{m}-1-k}}_{\mathbb P(n=k)} \nonumber \\ 
&\stackrel{(d)}{=}\sum_{k=0}^{N_{m}-1}\left(\int_{0}^{\infty}\frac{1}{1+sP_{t}v_{1}^{-\alpha}}f_{V_{1}}(v_{1})dv_{1}\right)^{k}\times {{N_{m}-1}\choose{k}}(\mathbb P_{S,S})^{k}(1-\mathbb P_{S,S})^{N_{m}-1-k} \nonumber \\ 
&=\sum_{k=0}^{N_{m}-1}{{N_{m}-1}\choose{k}}\left(\int_{0}^{\infty}\frac{1}{1+sP_{t}z_{1}^{-\alpha}}f_{V_{1}}(v_{1})dv_{1}\mathbb P_{S,S}\right)^{k}\times(1-\mathbb P_{S,S})^{N_{m}-1-k} \nonumber \\
&\stackrel{(e)}{=}\left(\int_{0}^{\infty}\frac{1}{1+sP_{t}v_{1}^{-\alpha}}f_{V_{1}}(v_{1})dv_{1}\mathbb P_{S,S}+1-\mathbb P_{S,S}\right)^{N_{m}-1},
\end{align}
}%
where (a) is the definition of Laplace transformation, (b) follows from assumption of Rayleigh fading channel and the channel gain $h_{yj}\sim \exp(1)$, and the transformation in (c) is due to the fact the locations of devices within the cluster are i.i.d. variables. We substitute $v_{1}$ with $||y-y_{j}||$ and change the Cartesian to the polar coordinates in (d). Since both $y$ and $y_{j}$ are i.i.d., and follow zero mean complex Gaussian distribution with variance $\sigma^{2}$,  we have the PDF of $v_{1}$: $f_{V_{1}}(v_{1})=f_{S}(v_{1};2\sigma^{2})$, $v_{1}\geq0$. The change in (e) is based upon the binomial expansion of a power of a sum \cite{harris1998handbook}. 
\vspace{-0.5cm}


\subsection{Proof of Lemma \ref{lemmasecond}}
\vspace{-0.1cm}
\label{Appendix C}
We can conduct Laplace transform for the inter-cluster interference as
{\small
\begin{align}
\label{inter1}
\mathcal{L}^{\textrm{star}}_{j\text{-inter}}(s)&\stackrel{(a)}{=}\E\left[\exp\left(-s\sum_{x\in \Phi_{p}\setminus x_{r}} \sum_{y^{\prime} \in \mathcal{T}_{x}} P_{t}h_{yj}||x+y^{\prime}-x_{r}-y_{j}||^{-\alpha}\right)\right]  \nonumber \\
&=\E_{ \Phi_{p}}\left[\prod_{x\in \Phi_{p}\setminus x_{r}}\E_{\mathcal{T}_{x}}\left[\prod_{y^{\prime}\in \mathcal{T}_{x}}\E_{h_{yj}}[\exp(-sP_{t}h_{yj}||x+y^{\prime}-x_{r}-y_{j}||^{-\alpha})]\right]\right] \nonumber \\
&\stackrel{(b)}{=}\E_{ \Phi_{p}}\left[\prod_{x\in \Phi_{p}\setminus x_{r}}\E_{\mathcal{T}_{x}}\left[\prod_{y^{\prime}\in \mathcal{T}_{x}}\frac{1}{1+sP_{t}||x+y^{\prime}-x_{r}-y_{j}||^{-\alpha}}\right]\right] \nonumber \\
&\stackrel{(c)}{=}\E_{ \Phi_{p}}\Bigg[\prod_{x\in \Phi_{p}\setminus x_{r}} \sum_{k=0}^{N_{m}}\left(\E_{\mathcal{T}_{x}}\left[\frac{1}{1+sP_{t}||x+y^{\prime}-x_{r}-y_{j}||^{-\alpha}}\right]\right)^{k}\times\nonumber\\ &\underbrace{{{N_{m}}\choose{k}}(\mathbb P_{S,S})^{k}(1-\mathbb P_{S,S})^{N_{m}-k}}_{\mathbb P(n=k)}\Bigg] \nonumber \\
&\stackrel{(d)}{=}\E_{ \Phi_{p}}\left[\prod_{x\in \Phi_{p}\setminus x_{r}} \sum_{k=0}^{N_{m}}\left(\int_{\R^{2}}\frac{f_{Z_{2}}(z_{2})dz_{2}}{1+sP_{t}||x+z_{2}||^{-\alpha}}\right)^{k}\!\!\times\!\! {{N_{m}}\choose{k}}(\mathbb P_{S,S})^{k}(1\!\!-\!\!\mathbb P_{S,S})^{N_{m}-k}\right] \nonumber \\
&\stackrel{(e)}{=}\exp\left(\!-\!\lambda_{p}\!\!\int_{\R^{2}}\!\!\left(1\!\!-\!\!\sum_{k=0}^{N_{m}}\left(\int_{\R^{2}}\!\!\frac{f_{Z_{2}}(z_{2})dz_{2}}{1\!\!+\!\!sP_{t}||x\!\!+\!\!z_{2}||^{-\alpha}}\right)^{k} 
\!\!\times\!\! {{N_{m}}\choose{k}}(\mathbb P_{S,S})^{k}(1\!\!-\!\!\mathbb P_{S,S})^{N_{m}\!-\!k}\right)\!dx\!\right) \nonumber \\
&\stackrel{(f)}{=}\exp\left(-\lambda_{p}\int_{\R^{2}}\left(1-\left(\int_{\R^{2}}\frac{f_{Z_{2}}(z_{2})dz_{2}\mathbb P_{S,S}}{1+sP_{t}||x+z_{2}||^{-\alpha}}+1-\mathbb P_{S,S}\right)^{N_{m}}\right)dx\right)\nonumber \\
&\stackrel{(g)}{=}\exp\left(-\lambda_{p} 2\pi \int_{0}^{\infty}\left(1-\left(\int_{0}^{\infty}\frac{f_{V_{2}}(v_{2}|t)dv_{2}\mathbb P_{S,S}}{1+sP_{t}v_{2}^{-\alpha}}+1-\mathbb P_{S,S}\right)^{N_{m}}\right)tdt\right),
\end{align}
}%
where $f_{V_{2}}(v_{2}|t)=f_{D}(v_{2},t;3\sigma^{2})$, $v_{2}\geq 0$. The changes in (a), (b), (c), and (f) follow the same reasons with (a), (b), (c) and (e) in (\ref{f251}). In (d), we assume $z_{2}=y^{\prime}-x_{r}-y_{i}$, where $y^{\prime}$, $x_{r}$, and $y_{1}$ are i.i.d. and follow zero mean complex Gaussian distribution with variance $\sigma^{2}$. Hence, $z_{2}$ follows zero mean complex Gaussian distribution with variance $3\sigma^{2}$. (e) can be explained by the probability generating functional (PGFL) of PPP \cite{chiu2013stochastic}. In (g), we substitute the variable $v_{2}$ with $||x+z_{2}||$ and convert the coordinates from Cartesian to polar.\vspace{-0.5cm}


\subsection{Proof of Lemma \ref{lemmathird}}
\vspace{-0.1cm}
We conduct the Laplace transform of the intra-cluster interference as 
\label{Appendix D}
{\small
\begin{align}
\mathcal{L}^{\textrm{chain}}_{j\text{-intra}}(s|s_{r})&\stackrel{(a)}{=}\E\left[\exp\left(-s\sum_{y\in \mathcal{T}_{x} \setminus y_{i}}P_{t}h_{yj}||x_{r}+y||^{-\alpha}\right)\right] \nonumber \\
&=\E_{\mathcal{T}_{x}}\left[\prod_{y\in \mathcal{T}_{x} \setminus y_{i}}\E_{h_{yj}}[\exp(-sP_{t}h_{yj}||x_{r}+y||^{-\alpha})]\right] \nonumber \\
&\stackrel{(b)}{=}\E_{ \mathcal{T}_{x}}\left[\prod_{y\in \mathcal{T}_{x} \setminus y_{j}}\E_{h_{yj}}\left[\frac{1}{1+sP_{t}||x_{r}+y||^{-\alpha}}\right]\right
] \nonumber \\
&\stackrel{(c)}{=}\sum_{k=0}^{N_{m}-1}\left(\int_{\R^{2}}\frac{f_{Y}(y)dy}{1+sP_{t}||x_{r}+y||^{-\alpha}}\right)^{k}\times \underbrace{{{N_{m}\!-\!1}\choose{k}}(\mathbb P_{S,S})^{k}(1\!-\!\mathbb P_{S,S})^{N_{m}-1-k}}_{\mathbb P(n=k)} \nonumber \\ 
&=\sum_{k=0}^{N_{m}\!-\!1}{{N_{m}\!-\!1}\choose{k}}\left(\int_{\R^{2}}\frac{1}{1\!+\!sP_{t}||x_{r}\!+\!y||^{-\alpha}}f_{Y}(y)dy\mathbb P_{S,S}\right)^{k}\times(1\!-\!\mathbb P_{S,S})^{N_{m}\!-\!1\!-\!k} \nonumber \\
&\stackrel{(d)}{=}\left(\int_{\R^{2}}\frac{1}{1+sP_{t}||x_{r}+y||^{-\alpha}}f_{Y}(y)dy\mathbb P_{S,S}+1-\mathbb P_{S,S}\right)^{N_{m}-1}
\nonumber\\
&\stackrel{(e)}{=}
\left(\int_{0}^{\infty}\frac{1}{1+sP_{t}v_{3}^{-\alpha}}f_{V_{3}}(v_{3}|s_{r})dv_{3}\mathbb P_{S,S}+1-\mathbb P_{S,S}\right)^{N_{m}-1},
\end{align}
}%
where $f_{R}(v_{3}|s_{r})=f_{D}(v_{3},s_{r};\sigma^{2})$. The changes in (a), (b), (c), and (d) follow the same reasons with (a), (b), (c) and (e) in (\ref{f251}). In (e), we change the variables by using $v_{3}=||x_{r}+y||$ and convert from Cartesian to polar coordinates.

\subsection{Proof of Lemma \ref{lemmafourth}}
\vspace{-0.1cm}
\label{Appendix E}
We conduct Laplace transform for the inter-cluster interference as 
{\small
\begin{align}
\label{f25}
\mathcal{L}^{\textrm{chain}}_{j\text{-inter}}(s)&\stackrel{(a)}{=}\E\left[\exp\left(-s\sum_{x\in \Phi_{p}\setminus x_{r}} \sum_{y^{\prime} \in \mathcal{T}_{x}} P_{t}h_{yj}||x+y^{\prime}||^{-\alpha}\right)\right]  \nonumber \\
&=\E_{ \Phi_{p}}\left[\prod_{x\in \Phi_{p}\setminus x_{r}}\E_{\mathcal{T}_{x}}\left[\prod_{y^{\prime}\in \mathcal{T}_{x}}\E_{h_{yj}}[\exp(-sP_{t}h_{yi}||x+y^{\prime}||^{-\alpha})]\right]\right] \nonumber \\
&\stackrel{(b)}{=}\E_{ \Phi_{p}}\left[\prod_{x\in \Phi_{p}\setminus x_{r}}\E_{\mathcal{T}_{x}}\left[\prod_{y^{\prime}\in \mathcal{T}_{x}}\frac{1}{1+sP_{t}||x+y^{\prime}||^{-\alpha}}\right]\right] \nonumber \\
&\stackrel{(c)}{=}\E_{ \Phi_{p}}\Bigg[\prod_{x\in \Phi_{p}\setminus x_{r}} \sum_{k=0}^{N_{m}}\left(\int_{\R^{2}}\frac{f_{Y}(y^{\prime})dy^{\prime}}{1\!+\!sP_{t}||x\!+\!y^{\prime}||^{-\alpha}}\right)^{k}\!\!\times\!\! \underbrace{{{N_{m}}\choose{k}}(\mathbb P_{S,S})^{k}(1\!-\!\mathbb P_{S,S})^{N_{m}-k}}_{\mathbb P(n=k)}\Bigg] \nonumber \\
&\stackrel{(d)}{=}\exp\Bigg(-\lambda_{p}\int_{\R^{2}}\Bigg(1-\sum_{k=0}^{N_{m}}\left(\int_{\R^{2}}\frac{1}{1+sP_{t}||x+y^{\prime}||^{-\alpha}}f_{Y}(y^{\prime})dy^{\prime}\right)^{k} \nonumber \\
&\times {{N_{m}}\choose{k}}(\mathbb P_{S,S})^{k}(1-\mathbb P_{S,S})^{N_{m}-k}\Bigg)dx\Bigg) \nonumber \\
&\stackrel{(e)}{=}\exp\left(-\lambda_{p}\int_{\R^{2}}\!\!\left(1\!-\!\left(\int_{\R^{2}}\!\!\frac{1}{1\!+\!sP_{t}||x\!+\!y^{\prime}||^{-\alpha}}f_{Y}(y^{\prime})dy^{\prime}\mathbb P_{S,S}\!+\!1\!-\!\mathbb P_{S,S}\right)^{N_{m}}\right)dx\right)\nonumber \\
&\stackrel{(f)}{=}\exp\left(-\lambda_{p} 2\pi \int_{0}^{\infty}\!\!\left(1\!-\!\left(\int_{0}^{\infty}\!\!\frac{1}{1\!+\!sP_{t}v_{4}^{-\alpha}}f_{V_{4}}(v_{4}|t)dv_{4}\mathbb P_{S,S}\!+\!1\!-\!\mathbb P_{S,S}\right)^{N_{m}}\right)tdt\right),
\end{align}
}%
where $f_{V_{4}}(v_{4}|t)=f_{D}(v_{4},t;\sigma^{2})$. The changes in (a), (b), (c) (d), and (e) follow the same reasons with (a), (b), (c), (d), and (e) in (\ref{inter1}). In (f), we change the variables by using $v_{4}=||x+y^{\prime}||$ and convert from Cartesian to polar coordinates.

\subsection{Proof of Theorem \ref{lemma4}}
\vspace{-0.1cm}
\label{AppendixOutageChainProof}
To derive the outage probability of chain motifs, we take into account the correlation in the interference $I_{t_{1}}(j)$ and $I_{t_{2}}(k)$ in PCP networks, $t_{1}\!\neq\!t_{2}, j\!\neq\!k$, where $I_{t_{1}}(j)$ and $I_{t_{2}}(k)$ represent the interference at time instant $t_{1}$ received by device $j$ and the interference at time instant $t_{2}$ received by device $k$, respectively.
The joint Laplace transform can be expressed as 
{\small
	\begin{align}
\label{jointLaplace}
\mathcal{L}(s_{1},s_{2})&=\E\exp\left[-s_{1}\sum_{x\in \Phi_{j} \backslash i}P_{t}h_{xj}(t_{1})(||x||)^{-\alpha}-s_{2}\sum_{y\in\Phi_{k}\backslash j}P_{t}h_{yk}(t_{2})(||y-(x_{r}+y_{k})||)^{-\alpha}\right] \nonumber \\
=&\E\prod_{x\in \Phi}\exp \left(-s_{1}\mathbbm{1}(x\in\Phi_{j}\backslash i )P_{t}h_{xj}(t_{1})(||x||)^{-\alpha}-s_{2}\mathbbm{1}(x\in\Phi_{k} \backslash j)P_{t}h_{xk}(t_{2})(||x-(x_{r}+y_{k})||)^{-\alpha} \right) \nonumber \\
\stackrel{(a)}{=}&\E\!\prod_{x\in \Phi}\!\left[1\!-\!p\!+\!p \exp\left(\!-s_{1}P_{t}h_{xj}(t_{1})(||x||)^{-\alpha}\right)\right]\!\!\left[1\!-\!p\!+\!p\exp\left(\!-s_{2}P_{t}h_{xk}(t_{2})(||x\!-\!(x_{r}\!+\!y_{k})||)^{-\alpha}\right)\right] \nonumber \\
\stackrel{(b)}{=}&\E\prod_{x\in \Phi}\left[1-p+p
\mathcal{L}_{h}(s_{1}P_{t}(||x||)^{-\alpha})\right] \left[1-p+p\mathcal{L}_{h}(s_{2}P_{t}(||x-(x_{r}+y_{k})||)^{-\alpha})\right] \nonumber \\
\stackrel{(c)}{=}&\exp\Bigg(-\lambda_{p}\int_{\R^{2}}\Big[1-\Big(\int_{\R^{2}}\left[1-p+p
\mathcal{L}_{h}(s_{1}P_{t}(||x+y||)^{-\alpha})\right] \nonumber \\  &\left[1-p+p\mathcal{L}_{h}(s_{2}P_{t}(||x+y-(x_{r}+y_{k})||)^{-\alpha})\right]  \frac{1}{2\pi\sigma^{2}} \exp\left(\frac{-||y||^{2}}{2\sigma^{2}}\right)dy
    \Big)\Big]dx\Bigg).
\end{align}
}%
In (a), we take into account the possibility of an arbitrarily selected node being a transmitter in the network, where $p=\frac{N_{m} \times  \mathbb P_{S,S}}{N}$. 
Since fading is independent across time and space, we move the expectation with respect to fading inside in (b). 
The changes in (c) follow the PGFL of the Neyman-Scott cluster process \cite{chiu2013stochastic}. 
Based on the joint Laplace transform, the outage probability of chain motifs can be calculated as
{\small
	\begin{align}
P^{\textrm{chain}}_{\textrm{outage}}(\delta_{\textrm{th}})&=1-\mathbb{P}[\delta_{ij}\geq \delta_{\textrm{th}}, \delta_{jk}\geq \delta_{\textrm{th}}]
\nonumber \\ 
&=1-\mathbb{P}\left[\frac{P_{t}h_{ij}d_{ij}^{-\alpha}}{I^{x_{r}}_{j-\textrm{intra}}+I^{x_{r}}_{j-\textrm{inter}}}\geq \delta_{\textrm{th}}, \frac{P_{t}h_{jk}d_{jk}^{-\alpha}}{I^{x_{r}}_{k-\textrm{intra}}+I^{x_{r}}_{k-\textrm{inter}}}\geq \delta_{\textrm{th}}   \right]  \nonumber \\
&= 1- \mathbb{P}\left[h_{ij}\geq \left( \frac{(I^{x_{r}}_{j-\textrm{intra}}+I^{x_{r}}_{j-\textrm{inter}})\delta_{\textrm{th}}}{P_{t}d_{ij}^{-\alpha}} \right) ,h_{jk}\geq \left( \frac{(I^{x_{r}}_{k-\textrm{intra}}+I^{x_{r}}_{k-\textrm{inter}})\delta_{\textrm{th}}}{P_{t}d_{jk}^{-\alpha}} \right) \right]  \nonumber \\
&\stackrel{(a)}{=} 1- \E\left[ \exp\left(\frac{-(I^{x_{r}}_{j-\textrm{intra}}+I^{x_{r}}_{j-\textrm{inter}})\delta_{\textrm{th}}}{P_{t}d_{ij}^{-\alpha}}\right)   \exp\left( \frac{-(I^{x_{r}}_{k-\textrm{intra}}+I^{x_{r}}_{k-\textrm{inter}})\delta_{\textrm{th}}}{P_{t}d_{jk}^{-\alpha}} \right)  \right] \nonumber \\
& = 1 - \E \left[ \exp\left(\frac{-(I^{x_{r}}_{j-\textrm{intra}}+I^{x_{r}}_{j-\textrm{inter}})\delta_{\textrm{th}}}{P_{t}d_{ij}^{-\alpha}}+ \frac{-(I^{x_{r}}_{k-\textrm{intra}}+I^{x_{r}}_{k-\textrm{inter}})\delta_{\textrm{th}}}{P_{t}d_{jk}^{-\alpha}}    \right) \right]  \nonumber \\ 
& \stackrel{(b)}{=}  1 - \mathcal{L}\left(\frac{\delta_{\textrm{th}}}{P_{t}d_{ij}^{-\alpha}}, \frac{\delta_{\textrm{th}}}{P_{t}d_{jk}^{-\alpha}}\right) = 1- \mathcal{L}\left(\frac{\delta_{\textrm{th}}}{P_{t}||x_{r}+y_{i}||^{-\alpha}}, \frac{\delta_{\textrm{th}}}{P_{t}||x_{r}+y_{k}||^{-\alpha}}\right),
\end{align}
}%
step in (a) stems from the fact that the channel gain of Rayleigh fading channel follows an exponential distribution, and in (b), we use the joint Laplace transform from (\ref{jointLaplace}).
After replacing $s_{1}$, $s_{2}$ with $\frac{\delta_{\textrm{th}}}{P_{t}||x_{r}+y_{i}||^{-\alpha}}$ and $\frac{\delta_{\textrm{th}}}{P_{t}||x_{r}+y_{k}||^{-\alpha}}$ and considering the distribution of $x_{r}+y_{i}$ and $x_{r}+y_{i}$, we can have the results in Theorem \ref{lemma4}.
\vspace{-0.1cm}

\def\baselinestretch{1.08}
\bibliographystyle{IEEEtran}

\end{document}